\documentclass[a4paper,11pt]{article}
\usepackage{jheppub}

\pdfoutput=1

\usepackage{hyperref}
\usepackage{graphicx}
\usepackage[utf8]{inputenc}
\usepackage{xspace}
\usepackage{amsmath,amssymb}
\usepackage{bbm}
\usepackage[usenames,dvipsnames]{xcolor}
\usepackage{booktabs}
\usepackage{subcaption}
\usepackage{mathtools}
\usepackage[ruled,norelsize]{algorithm2e}
\usepackage{empheq}
\usepackage{cancel}
\usepackage{multirow}
\usepackage{slashed}

\makeatletter
\g@addto@macro\bfseries{\boldmath}
\makeatother

\newcommand{\Kcmw}{K_\text{\textsc{cmw}}}

\newcommand{\order}[1]{{\cal O}\left(#1\right)}
\newcommand{\abar}{\bar{\alpha}}
\newcommand{\Bbar}{\bar{B}}

\newcommand{\as}{\alpha_s}

\newcommand{\sd}{d}

\newcommand{\cP}{\mathcal{P}}
\newcommand{\cR}{\mathcal{R}}
\newcommand{\MSbar}{\ensuremath{\overline{\text{MS}}}\xspace}

\usepackage[normalem]{ulem}
\DeclareFontFamily{U}{rcjhbltx}{}
\DeclareFontShape{U}{rcjhbltx}{m}{n}{<->rcjhbltx}{}

\newcommand{\rd}{\text{d}}

\title{A collinear shower algorithm for NSL non-singlet
    fragmentation}

\author[a]{Melissa van Beekveld,}
\author[b]{Mrinal Dasgupta,}
\author[c]{Basem Kamal El-Menoufi,}
\author[d]{Jack Helliwell,}
\author[e]{Pier Francesco Monni,}
\author[d,f]{Gavin P.\ Salam}

\affiliation[a]{Nikhef, Theory Group, Science Park 105, 1098 XG, Amsterdam, The Netherlands}
\affiliation[b]{Department of Physics \& Astronomy, University of
	Manchester, Manchester M13 9PL, United Kingdom}
\affiliation[c]{School of Physics and Astronomy, Monash University, Wellington Road, Clayton, VIC-3800, Australia}
\affiliation[d]{Rudolf Peierls Centre for Theoretical Physics, Clarendon
	Laboratory, Parks Road, University of Oxford, Oxford OX1 3PU, UK}
\affiliation[e]{CERN, Theoretical Physics Department, CH-1211 Geneva
  23, Switzerland}
\affiliation[f]{All Souls College, University of Oxford, Oxford OX1 4AL}

\emailAdd{mbeekvel@nikhef.nl}
\emailAdd{mrinal.dasgupta@manchester.ac.uk}
\emailAdd{basem.el-menoufi@monash.edu}
\emailAdd{jack.helliwell@physics.ox.ac.uk}
\emailAdd{pier.monni@cern.ch}
\emailAdd{gavin.salam@physics.ox.ac.uk}

\preprint{CERN-TH-2024-125, OUTP-24-05P, Nikhef 2024-013}

\abstract{%
  We formulate a collinear partonic shower algorithm that achieves
  next-to-single-logarithmic (NSL, $\as^n L^{n-1}$) accuracy for
  collinear-sensitive non-singlet fragmentation observables.
  This entails the development of an algorithm for nesting 
  triple-collinear splitting functions.
  It also involves the inclusion of the one-loop double-collinear
  corrections, through a $z$-dependent NLO-accurate effective $1\to 2$
  branching probability, using a formula that can be applied more
  generally also to future full showers with $1\to3$
  splitting kernels.
  The specific NLO branching probability is calculated in two ways,
  one based on slicing, the other using a subtraction approach based
  on recent analytical calculations.
  We close with
  demonstrations of the shower's accuracy for non-singlet partonic
  fragmentation functions and the energy spectrum of small-$R$ quark
  jets. This work represents an important conceptual step towards general NNLL accuracy in parton showers.
}

\begin{document}

\setlength{\parskip}{0pt}
\maketitle
\flushbottom

\section{Introduction}
\label{sec:intro}
Recent years have seen extensive work by multiple groups towards the
development of parton showers that achieve next-to-leading logarithmic
(NLL)
accuracy~\cite{Dasgupta:2018nvj,Dasgupta:2020fwr,Hamilton:2020rcu,Karlberg:2021kwr,Hamilton:2021dyz,vanBeekveld:2022zhl,vanBeekveld:2022ukn,vanBeekveld:2023chs,Forshaw:2020wrq,Nagy:2020dvz,Nagy:2020rmk,Herren:2022jej,Assi:2023rbu,Preuss:2024vyu,Hoche:2024dee}.
Broadly speaking, NLL accuracy implies control of terms $\as^n L^n$,
where $\as$ is the strong coupling and $L$ is the logarithm of the
ratio of any pair of disparate scales.

One of the next frontiers is to develop NNLL accurate showers, with control of
terms $\as^n L^{n-1}$.
A critical ingredient is the incorporation of splitting functions
beyond first order in the 
coupling~\cite{PhysRevD.36.61,PhysRevD.39.156,KATO199167,Kato:1991fs,Jadach:2011kc,Hartgring:2013jma,Jadach:2013dfd,Jadach:2016zgk,Li:2016yez,Hoche:2017iem,Hoche:2017hno,Dulat:2018vuy,Campbell:2021svd,Gellersen:2021eci,FerrarioRavasio:2023kyg,vanBeekveld:2024wws}.
However, on its own, higher accuracy of the splitting function is not
sufficient to achieve higher logarithmic accuracy.
Concentrating on final-state showers, one significant recent step has
been the inclusion of double-soft
corrections~\cite{FerrarioRavasio:2023kyg} in such a way as to achieve
$\as^n L^{n-1}$ (next-to-single logarithmic --- NSL)
accuracy~\cite{Banfi:2021owj,Banfi:2021xzn,Becher:2023vrh} 
for observables like the distribution of energy flow in any given
limited angular region, and $\as^n L^{2n-2}$ for subjet
multiplicities~\cite{Medves:2022uii,Medves:2022ccw}.
Another has been the development of an understanding of the
connections between the soft-collinear, large-angle soft and
triple-collinear regions.
Together these advances have enabled the PanScales parton shower
project to achieve NNLL accuracy for global and non-global event shape
observables in $Z \to q\bar q$ and
  $H \to gg$ processes~\cite{vanBeekveld:2024wws} (for corresponding
  calculations at NNLL and beyond see e.g.\ Refs.~\cite{%
    deFlorian:2004mp,
    Becher:2008cf,
    Abbate:2010xh,
    Chien:2010kc,
    Monni:2011gb, 
    Becher:2012qc, 
    Hoang:2014wka, 
    Banfi:2014sua, 
    Banfi:2016zlc, 
    Frye:2016okc, 
    Frye:2016aiz, 
    Tulipant:2017ybb,
    Moult:2018jzp, 
    Bell:2018gce, 
    Banfi:2018mcq, 
    Procura:2018zpn, 
    Arpino:2019ozn, 
    Bauer:2020npd, 
    Kardos:2020gty, 
    Anderle:2020mxj, 
    Duhr:2022yyp,
    Dasgupta:2022fim
    }).
One advance that is still needed for general NNLL accuracy in
parton showers is the full treatment of the triple-collinear
region.
Arguably this is the last major step for general
final-state NNLL accuracy at leading colour aside from NNLL spin
correlations.
It is required, for example, for
$\as^n L^{n-1}$ accuracy for phenomenologically important quantities
such as fragmentation functions and many jet substructure observables.
Observables of this class are common in collider physics and have been
extensively studied in the literature (see
e.g.~\cite{Gribov:1972ri,Dokshitzer:1977sg,Altarelli:1977zs,Furmanski:1980cm,Curci:1980uw,Jain:2011xz,Alioli:2013hba,Chang:2013rca,Ritzmann:2014mka,Dasgupta:2014yra,Banfi:2015pju,Dasgupta:2016bnd,Kang:2016mcy,Dixon:2019uzg,Chen:2020adz,Chen:2021gdk,Dasgupta:2021hbh,Li:2021zcf,Dasgupta:2022fim,Chen:2022jhb,Chen:2022muj,Liu:2022wop,Chen:2023zlx,vanBeekveld:2023lsa,Lee:2023npz,Lee:2024esz,Chen:2024nyc}).
Its inclusion in showers would also open the door to potential new
methods for merging parton showers with NNLO calculations (see e.g.\
Refs.~\cite{Hamilton:2013fea,Alioli:2013hqa,Hoche:2014uhw,Monni:2019whf,Monni:2020nks,Alioli:2021qbf,Campbell:2021svd}
for existing methods), because it would ensure that parton showers
reproduce all divergences that appear up to and including order $\as^2$.
Several groups have explored the triple collinear region for parton
showers, both in the
final and initial
state~\cite{Jadach:2011kc,Jadach:2013dfd,Jadach:2016zgk,Li:2016yez,Hoche:2017iem,Hoche:2017hno},
though so far there have been only limited tests of their implications
for logarithmic accuracy~\cite{Jadach:2013dfd}.

In this work, we take an exploratory approach to the problem of
achieving $\as^n L^{n-1}$ in the collinear final-state regime, with the ultimate
perspective of porting the lessons that we learn into the context of
full showers, notably those being developed in the PanScales project.
In Section~\ref{sec:form-toy-show}, we establish how to include
 the Abelian part of $q \to q g g$ branching~\cite{Campbell:1997hg,Catani:1998nv} in a
positive-definite, unit-weight shower formalism.
In Section~\ref{sec:virtual-general}, we identify how to treat virtual
corrections~\cite{Bern:1994zx,Bern:1995ix,Kosower:1999xi,Kosower:1999rx,Sborlini:2013jba}, with a core formula whose applicability is wider than
parton-shower logarithmic accuracy and that connects with the work of
Refs.~\cite{Hartgring:2013jma,Li:2016yez,Campbell:2021svd}.
Then in Section~\ref{sec:K-determination} we use that formula to
evaluate the resulting NLO-accurate branching probabilities
specifically for $q \to qg$ splitting, in both slicing and subtraction
approaches, the latter building on the work of
Refs.~\cite{Dasgupta:2021hbh,vanBeekveld:2023lsa}.
Together, these ingredients are sufficient for us to demonstrate, in
Section~\ref{sec:log-tests}, $\as^n L^{n-1}$ accuracy for two distinct
observables, in the limit of a large number of colours: the non-singlet fragmentation function \cite{Furmanski:1980cm,Curci:1980uw} and the small-$R$ inclusive non-singlet
quark-jet momentum distribution, with comparisons to calculations from
Ref.~\cite{vanBeekveld:2024jnx}.\footnote{A conjecture for the two-loop anomalous dimension for small radius jet evolution was previously made in Ref.~\cite{Kang:2016mcy}.}
We conclude in Section~\ref{sec:conclusions} and also provide
additional technical appendices.

\section{Formulation of a non-singlet final-state collinear shower}
\label{sec:form-toy-show}

Let us start by formulating a standard strongly ordered non-singlet
collinear shower.
It involves an ordering scale $v$ and the iteration of $1\to2$
splitting steps, each at a successively smaller values of $v$.
It is useful to introduce a generic real branching probability
\begin{equation}
  \label{eq:real-1to2}
  d\cR_{1\to2}(v_i, z_i, \phi_i | v_{i-1})
  = 
  \frac{dv_i}{v_i} dz_i \frac{d\phi_i}{2\pi}
  \frac{\as(\mu)}{\pi}
  P_{qq}(z_i)
  \Theta(v_i < v_{i-1})
  \,,
\end{equation}
where $v_i$ has dimensions of energy, e.g.\ a transverse momentum and
we have included the ordering condition relative to a previous
emission at scale $v_{i-1}$. The renormalisation scale $\mu$ is to be
taken of the order of the transverse momentum of the emitted gluon,
and the splitting function is given by
\begin{equation}
P_{qq}(z) = C_F\, p_{qq}(z), \quad \text{with} \quad p_{qq}(z) = \frac{1+z^2}{1-z}\,.
\end{equation}
Then the distribution of branching $i$ is given by
\begin{equation}
  \label{eq:full-1to2}
  d\cP_i
  = d\cR_{1\to2}(v_i,z_i,\phi_i \,|\, v_{i-1}) \, \Delta(v_{i-1},v_i)\, \,,
\end{equation}
where the Sudakov form factor $\Delta(v_{i-1},v_i)$ is specified
through unitarity,
\begin{equation}
  \label{eq:Sudakov-1to2}
  \Delta(v_{i-1},v_i) = \exp\left[-\int d\cR_{1\to2}(v,z,\phi\,|\,v_{i-1}) \,\Theta(v > v_i)\right].
\end{equation}
This is a standard formulation that guarantees single-logarithmic
accuracy for collinear observables such as non-singlet fragmentation
functions.

\begin{figure}
  \centering
  \includegraphics[width=0.7\textwidth]{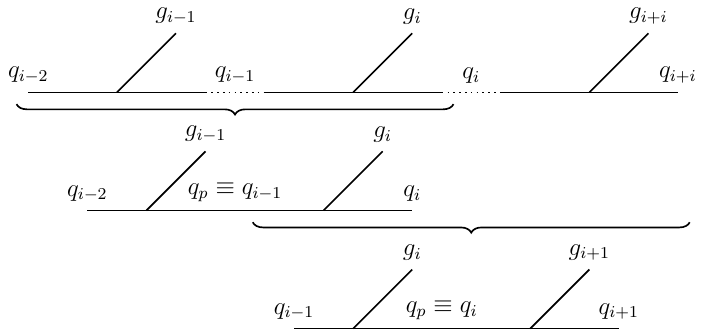}
  \caption{Cartoon illustration of how the shower builds up the $1\to3$
    splitting functions from pairwise combinations of $1 \to 2$
    splittings. 
    The top line shows how the shower builds up a sequence of $1\to2$ splittings, while the second and third lines show how these are grouped and calculated using iterated $1\to3$ splitting functions.
    This cartoon ignores subtleties around
    	unordered emissions and the freedom to choose the parent quark $q_p$.
    	These subtleties are further discussed in the text and Appendix~\ref{sec:parent-finding}.
    }
  \label{fig:iterated-1to2}
\end{figure}

To achieve next-to-single logarithmic accuracy, we will need to modify the real branching probability so as to reproduce the full $1\to3$ matrix elements
(and phase space)~\cite{Campbell:1997hg,Catani:1998nv}.
We will also need to adjust the normalisation of splittings in the
strongly ordered $1 \to 2$ limit so as to take into account the
correct one-loop corrections to $1\to 2$
splittings~\cite{Bern:1994zx,Bern:1995ix,Kosower:1999xi,Kosower:1999rx,Sborlini:2013jba}. 

Let us first discuss how to incorporate the $1\to 3$ splitting matrix
element.
We will concentrate on the abelian-like $C_F^2$, $q\to qgg$
contribution, as relevant for the non-singlet case.\footnote{For the
  full evaluation of the non-singlet fragmentation functions, in
  addition to $q \to qgg$ one also needs the non-trivial
  ($C_F(2 C_F -C_A)$) part of the $q \to q q\bar q$, with identical
  flavours.
  In our numerical tests in Section~\ref{sec:log-tests}, we will work
  with colour factors chosen so that $C_A=2 C_F$, so that we can
  ignore this channel.}
Our approach will iterate $1 \to 2$ splittings, such that each step
takes into account the kinematics and matrix element used to generate
the prior $1\to 2$ splitting and includes corrections such that the
pair of splittings together has the correct $1\to 3$ splitting matrix element.
This is illustrated in Fig.~\ref{fig:iterated-1to2}. 
To be more specific, we concentrate on one specific pair of $1\to2$
splittings, which we label as
\begin{equation}
  \label{eq:notation-two-pairs}
  q_{i-2} \to g_{i-1} \, q_{i-1} \to g_{i-1} \, g_i \, q_i\,.
\end{equation}
Let us assume that the $q_{i-2} \to g_{i-1} \, q_{i-1}$
splitting has already been generated and was strongly ordered with
respect to any prior splittings.\footnote{At NSL accuracy, one need
  not worry about more than two consecutive commensurate-angle
  $1 \to 2$ splittings. }
We will write the real splitting probability as the ratio of the full
$1\to3$ splitting and phase space to the prior $1\to2$ splitting and
phase space.
First, for definiteness, it is useful to introduce the real phase space and matrix element for $1\to 3$ splitting,
\begin{align}
  \label{eq:1to3-fact}
  \left(\frac{\as}{\pi}\right)^2
  \left(
    \frac{\rd v_p}{v_p} \,
    \frac{\rd \phi_p}{2\pi}\,
    \rd z_p
  \right)
  \left(
     \frac{\rd v_i}{v_i}\,
    \frac{\rd \phi_i}{2\pi}\,
    \rd z_i
  \right)\,
  p_{1\to 3}(v_{p}, v_i, z_{p}, z_{i}, \phi_i - \phi_p)\,.
\end{align}
We have introduced the shorthand $p=i-1$,\footnote{Although it may be helpful to think of $p$ as corresponding to $i-1$ as in
  Fig.~\ref{fig:iterated-1to2}, this is not always the choice we make, as discussed in Appendix~\ref{sec:parent-finding}.
} and we have
\begin{align}
  \label{eq:p1to3}
  p_{1\to 3}(v_{p}, v_i, z_{p}, z_{i}, \phi)
  = z_{p}\, x_{g_{p}} x_{g_{i}} x_{q_{i}}
  \frac{E^4\,\theta_{g_{p}q_{p}}^2 \theta_{q_i g_i}^2}{s_{g_{p}g_i q_i}^2} \,
  C_F^2 \langle {\hat P}^{\text{(ab)}}_{q\to g_{p} g_i q_i}\rangle \,,
\end{align}
where $E$ is the energy of the three parton system and
$\langle {\hat P}^{\text{(ab)}}_{q\to g_{p} g_i q_i}\rangle$
corresponds to Eq.~(33) of Ref.~\cite{Catani:1998nv} (without any
$1/2!$ symmetry factor). The energy fractions and invariant mass of
the three particles involved in the triple-collinear spitting are
parameterised as
\begin{subequations}
	\label{eq:mappingxtoz}
  \begin{align}
    x_{g_p} &= 1-z_p \,, \\
    x_{g_i} &= z_p(1-z_i) \,, \\
    x_{q_i} &= z_p z_i \,,\\
    s_{ijk} &= x_i x_j \theta_{ij}^2 E^2 + \{i\leftrightarrow k\} + \{j\leftrightarrow k\}\,.
  \end{align}
\end{subequations}
Now we are in a position to write the probability for splitting $i$
given a previous splitting $p$,
\begin{equation}
  \label{eq:real-1to3}
  d\cR_{2\to3}(v_i,z_i,\phi_i \,|\, p)
  =
  \frac{\as^\text{eff}}{\pi}
  \left(
     \frac{\rd v_i}{v_i}\,
    \frac{\rd \phi_i}{2\pi}\,
    \rd z_i
  \right)\,
  \frac{p_{1\to 3}(v_{p}, v_i, z_{p}, z_{i}, \phi_i -
    \phi_p)}{P_{qq}(z_p)}
  \Theta(v_{g_i q_i} < v_{g_{p}q_i})\,,
\end{equation}
where we return to the definition and functional dependence of the
effective strength of emission, $\as^\text{eff}$, below.
An important consideration in Eq.~(\ref{eq:real-1to3}) is the
replacement of the condition $v_i < v_{i-1}$ in
Eq.~(\ref{eq:real-1to2}) with the condition
$v_{g_i q_i} < v_{g_{p} q_i}$.
Here $v_{bc}$ is a kinematic variable defined in terms of the momenta
of particles $b$ and $c$ (for example a relative transverse momentum),
chosen so as to coincide with the ordering variable $v$ for a
splitting $a \to bc$.
Thus $v_{g_i q_i}$ will coincide with $v_i$, but crucially
$v_{g_{p} q_i}$ differs from $v_{g_{p} q_p} \equiv v_p$, because $q_i$
will in general have a different momentum from $q_p$.
A key relevant feature of the $v_{g_i q_i} < v_{g_{p} q_i}$ condition is that it
is symmetric between the two gluons and therefore exactly accounts for
the $1/2!$ symmetry factor that needs to be associated with the
integral over the $1 \to 3$ phase space.\footnote{In this context it
  would be interesting to explore the connections with the sectoring
  approach of Ref.~\cite{Campbell:2021svd}.}
Note that $v_{g_{p} q_i}$ is known only after the generation of
splitting $i$ and that it can be either larger or smaller than $v_p$.
Therefore, in the above approach, one must not additionally impose a
$v_i < v_p$ condition.
Instead, we start from $v_i = X v_p$, where $X$ is large
enough to ensure that the vast majority of the
$v_{g_i q_i} < v_{g_{p} q_i}$ phase space is covered.
Finally, note that the Sudakov factor in Eq.~(\ref{eq:Sudakov-1to2})
also needs to be modified to use Eq.~(\ref{eq:real-1to3}) in its
integrand.

Now we turn to $\as^\text{eff}$.
Assuming that $v_i$ is a transverse-momentum like variable, then we
can write
\begin{equation}
  \label{eq:as-eff}
  \as^\text{eff} =
  \as(v_i) \left[
    1 +  \frac{\as(v_i)}{2\pi}K(z_i)
  \right] 
\end{equation}
where $K(z_i)$ generalises the well-known
$\Kcmw$~\cite{Catani:1990rr,Banfi:2018mcq,Catani:2019rvy} away from
the soft limit.
Its determination is discussed in
sections~\ref{sec:virtual-general} and \ref{sec:K-determination}.
As we shall see in those sections, $K(z_i)$ diverges logarithmically
when the quark becomes soft ($z_i \to 0$), which could cause
$\as^\text{eff}$ to become negative.
In practice therefore, we replace $\frac{\as(v_i)}{2\pi}K(z_i) \to
\tanh \left(\frac{\as(v_i)}{2\pi}K(z_i)\right)$.
A similar choice was made in
Refs.~\cite{FerrarioRavasio:2023kyg,vanBeekveld:2024wws}, which is
  allowed because this replacement generates only additional terms
  beyond the nominal $\alpha_s^n L^{n-1}$ logarithmic accuracy.

Next, we set out our choice of ordering variable.
We use
\begin{equation}
  \label{eq:vi-definition}
  v_{g_iq_i} = \min(E_{g_i}, E_{q_i}) \theta_{g_i q_i}\,,
\end{equation}
which coincides with transverse momentum in the soft-gluon limit.
Away from that limit one could have imagined a range of different
choices, but not all are equally straightforward to implement within
the above scheme.\footnote{In particular, we also investigated
  $v_{g_iq_i} = E_{g_i} \theta_{g_i q_i}$, but found that this had
    problems with boundedness of Eq.~(\ref{eq:real-1to3}) in the limit
    where the $p$ branching causes the quark $q_p$ to become soft and
    the $i$ branching causes the quark $q_i$ to end up close in angle to
    $g_p$.
  }
A final comment that we make concerns a subtlety with the above
picture connected with soft gluon emission.
Specifically, while the intent of Eq.~(\ref{eq:real-1to3}) is to
account for the $1 \to 3$ splitting function, there is a priori
freedom in terms of which previously emitted gluon should be taken as
the parent $g_p$.
Simply taking the parent gluon to be $p=i-1$ turns out to be incorrect
and infrared unsafe.
Instead we effectively take $p$ to be that among all prior
emissions that is closest in rapidity to $i$, with an additional
prescription to avoid double-counting of phase-space regions.
Appendix~\ref{sec:parent-finding} explains the issue and the details
of our approach.

\section{NLO-accurate inclusive emission probability}
\label{sec:virtual-general}

To understand how to calculate the $K(z)$ in Eq.~(\ref{eq:as-eff}), it
is helpful to consider a specific case, namely the emission of a
collinear gluon $g$ from an $e^+e^- \to q\bar q$ system.
We will do this in two parts: we will consider the actual NLO cross
section for producing a $q\bar q g$ system, inclusive over subsequent
branchings from that system;
separately we will consider the full shower expression for that gluon
emission up to NLO, Eq.~(\ref{eq:real-1to3}); then we will determine
$K(z)$ in Eq.~(\ref{eq:as-eff}) by equating the actual NLO
result and the shower's expansion to NLO.
We will see that the result that emerges is independent of 
the specific choice of system (here $e^+e^- \to q\bar q$) that we took as our starting
point, i.e.\ the process-dependent parts will cancel in $K(z)$.

We start with the first part.
We take a Born matrix element and phase space
$\abar B_{q\bar q g} d\Phi_{q\bar qg}$.
Here, $\abar = \as(\mu)/2\pi$, with $\mu$ corresponding to the
renormalisation scale, which should be taken of the order of the transverse momentum of the emitted gluon, as done explicitly in Eq.~\eqref{eq:as-eff}.
We use the convention of stripping out factors of $\abar$ from the various matrix elements.
The NLO-corrected version of $B_{q\bar q g}$ will be written
as $\bar B_{q\bar q g}$, with
\begin{equation}
  \label{eq:Bbar-qqg}
  \abar \Bbar_{q \bar q g} d\Phi_{q\bar qg}
  =
  \abar B_{q \bar q g} d\Phi_{q\bar qg} \left(
    1 + \abar \frac{V_{q\bar q g}}{B_{q\bar q g}} +
    \abar \int^{\tilde v_g}_0
    \frac{d\Phi_{q\bar q ij}}{d\Phi_{q\bar q g}}
    \frac{B_{q\bar q ij}}{B_{q \bar q g}}
  \right),
\end{equation}
where $V_{q\bar q g}$ is the one-loop correction to the $q\bar q g$
process, $B_{q\bar q ij}$ represents the matrix element for
one further branching.
The $\frac{d\Phi_{q\bar q ij}}{d\Phi_{q\bar q g}}$ factor represents
the phase space in the exact shower map for one branching given the
$q\bar q g$ starting point.
There is an implicit sum over the possible final state channels $ij =
\{gg, q\bar q\}$.
The upper limit $\tilde v_g$ in the integration is to be understood as
a shorthand for the ordering condition written in
Eq.~(\ref{eq:real-1to3}) --- or, more generally, the ordering
condition in any shower that generates the correct $1\to3$ splitting
functions.%
\footnote{Given an ordering condition, and given that the initial gluon
  $g$ is collinear, $i$ and $j$ will always be collinear and/or
  soft, as is required for the real phase space to be
    accurately generated with the triple collinear splitting
    function. Of course, the triple collinear splitting function does
    not generate the correct large-angle soft distribution, but as we
    shall see below, the contribution from this region cancels.
}
Eq.~(\ref{eq:Bbar-qqg}) is the way the MC@NLO or POWHEG
methods~\cite{Frixione:2002ik,Nason:2004rx} 
organise the NLO calculation given a Born $q\bar q g$
configuration.

Next consider the cross section that is obtained in the shower.
A crucial assumption is that the shower is NLO accurate for the $q\bar
q$ configuration, i.e.\ the starting point for the shower has a weight
\begin{equation}
  \label{eq:BBar-qq}
  \Bbar_{q \bar q} d\Phi_{q\bar q}
  =
 B_{q \bar q} d\Phi_{q\bar q} \left(
    1 + \abar \frac{V_{q\bar q}}{B_{q\bar q}} +
    \abar \int^{v_{\max}}_0
    \frac{d\Phi_{q\bar q g}}{d\Phi_{q\bar q}}
    \frac{B_{q\bar q g}}{B_{q \bar q}}
  \right),
\end{equation}
where the $0$ to ${v_{\max}}$ limits on the integral indicate the
range of allowed values of shower ordering variable $v_g$ associated
with the integration over the full $g$-emission phase space.
The shower weight to obtain a $q\bar q g$ system is equal to
\begin{equation}
  \label{eq:Sqqg}
  \abar S_{q\bar q g} \, d\Phi_{q\bar q g}
  =
  \Bbar_{q \bar q} \,d\Phi_{q\bar q}
  \times
  \Delta(v_{\max}, v_g)
  \times
  \abar(1 + \abar K(z)) 
  \frac{d\Phi_{q\bar q g}}{d\Phi_{q\bar q}}
  \frac{B_{q\bar q g}}{B_{q \bar q}}\,,
\end{equation}
where $K(z)$ is the shower NLO correction (to be determined) for the
emission of a gluon with longitudinal momentum fraction $1-z$ with
respect to its emitter.
The shower ordering variable at the point where the gluon is emitted
has a value $v_g$.
The Sudakov factor $\Delta(v_{\max}, v_g)$ is given by the unitary
counterpart of the real emission probability
\begin{subequations}
  \label{eq:Sudakov-qq}
  \begin{align}
    \Delta(v_{\max}, v_g)
    &= \exp \left[
      - \int^{v_{\max}}_{v_g}
      \abar(1 + \abar K(z')) 
      \frac{d\Phi_{q\bar q g'}}{d\Phi_{q\bar q}}
      \frac{B_{q\bar q g'}}{B_{q \bar q}}
      \right],
    \\
    &=
      1
      - \abar \int^{v_{\max}}_{v_g}
      \frac{d\Phi_{q\bar q g'}}{d\Phi_{q\bar q}}
      \frac{B_{q\bar q g'}}{B_{q \bar q}}
      + \order{\abar^2}\,.
  \end{align}
\end{subequations}
Putting together Eqs.~(\ref{eq:BBar-qq})--(\ref{eq:Sudakov-qq}), and writing
the result to relative order $\abar$ gives
\begin{equation}
  \label{eq:Sqqg-expanded}
  \abar S_{q\bar q g} \, d\Phi_{q\bar q g}
  =
  \abar B_{q\bar q g}  d\Phi_{q\bar q g} \left[
    1 + \abar \left(
      K(z)
      + \frac{V_{q\bar q}}{B_{q\bar q}}
      + \int^{v_g}_{0}
        \frac{d\Phi_{q\bar q g'}}{d\Phi_{q\bar q}}
        \frac{B_{q\bar q g'}}{B_{q \bar q}}
      \right)
      + \order{\abar^2}
  \right]\,,
\end{equation}
where we have combined the two integrals over $d\Phi_{q\bar q g}$.

Now we are ready to determine $K(z)$ by equating
$\abar \Bbar_{q\bar qg}$ and $\abar S_{q\bar q g}$ up to relative
order $\abar$.
This gives
\begin{equation}
  \label{eq:K-main}
  K(z) =
  \frac{V_{q\bar q g}}{B_{q\bar q g}}
  - \frac{V_{q\bar q}}{B_{q\bar q}}
  + \int^{\tilde v_g}_0
  \frac{d\Phi_{q\bar q ij}}{d\Phi_{q\bar q g}}
  \frac{B_{q\bar q ij}}{B_{q \bar q g}}
  - \int^{v_g}_{0}
  \frac{d\Phi_{q\bar q g'}}{d\Phi_{q\bar q}}
  \frac{B_{q\bar q g'}}{B_{q \bar q}}\,.
\end{equation}
There are important simplifications that can be made to Eq.~(\ref{eq:K-main}) thanks to the use of the collinear limit.
Assuming that the gluon is collinear to the quark,
Ref.~\cite{Sborlini:2013jba} has shown that
\begin{equation}
  \label{eq:Vdiff}
    \frac{V_{q\bar q g}}{B_{q\bar q g}}
  - \frac{V_{q\bar q}}{B_{q\bar q}}
  =
  \frac{P^{(1)}_{qq}(s_{qg}, z, \epsilon, \mu^2)}{P_{qq}(z,\epsilon)}\,,
\end{equation}
where $P^{(1)}_{qq}$ is a one-loop correction to the splitting
function, which is independent of the hard process.
We have made explicit that it depends on $s_{qg}$ (the $qg$ squared
invariant mass), the dimensional regularisation parameter $\epsilon$
and the renormalisation scale $\mu$.
The bare expression for $P^{(1)}_{qq}$ is given in
Eq.~(103) of Ref.~\cite{Sborlini:2013jba}.
The renormalised expressions that we use are given in Eq.~\eqref{eq:virt-corr}.
In dimensional regularisation, Eq.~(\ref{eq:Vdiff}) has a $1/\epsilon$
pole in the $C_F$ and $T_R n_f$ colour factors and a $1/\epsilon^2$
divergence in the $C_A$ colour factor.

Turning our attention now to the two real integrals of
Eq.~(\ref{eq:K-main}), several comments are in order.
Firstly, the $\tilde v_g$ and $v_g$ upper bounds act on different
phase spaces ($q\bar q g \to q\bar q ij$ and $q\bar q \to
  q\bar q g$ respectively) and so are not in general identical in the
two real integrals.
There is however one important region where they must coincide between
the integrals.
Consider, specifically the $ij = g_1 g_2$
channel in the left-hand integral.
Labelling $g_1$ as the smaller-angle gluon, it concerns the situation
where $\theta_{g_2 q} \gg \theta_{g_1 q}$, i.e.\ the angular
anti-strong-ordered situation.
Here the $\tilde v_g$ ordering condition on the $\{g_2,q\}$ phase
space in the left-hand integral coincides with the $v_g$
condition on the $\{g',q\}$ phase space in the right-hand integral.
Furthermore, in this region, the two integrals will have identical
integrands, because of the factorisation properties of the
$q\bar q g_1 g_2$ matrix element when one gluon ($g_1$) is collinear
and the other ($g_2$, over which we integrate) is soft and at much
larger angles.
Therefore this region cancels between the two integrals.
This is important, because it is the only region that carries
dependence on the hard process, specifically when $\theta_{g_2 q}$ is
of order $1$.
The cancellation ensures that the difference between the two integrals
is process independent and that one is therefore free to replace the
full matrix element with expressions involving just the universal
$1\to2$ and $1\to3$ splitting functions.
Together with the process-independence of Eq.~(\ref{eq:Vdiff}), this
ensures that $K(z)$ as a whole does not depend on the process.

An important observation that we make is that a formula analogous to Eq.~\eqref{eq:K-main} is valid for calculating the NLO-accurate emission probability generically for the $n^\text{th}$ emission in any parton shower that had NLO accuracy for all prior emissions $1 \ldots(n-1)$.
In particular, a similar form holds also for each of the
  nested soft and collinear gluon emissions that are the essence of
the NLO-accurate parton branching kernel as needed for $\as^n L^{n-1}$
accuracy.%
\footnote{Extending all of the ingredients beyond the collinear
  limit, such a formula is consistent also with the approach used in
  Ref.~\cite{FerrarioRavasio:2023kyg}.
    One way of understanding that approach is to note that all terms
    in Eq.~(\ref{eq:K-main}) are independent of the soft gluon
    rapidity, except the first real integral.
    Exploiting the fact that $K$ should agree with $\Kcmw$ in
    the soft-collinear limit, it is then straightforward to see that
    $K - \Kcmw$ is given by the difference of that first real
    term between the situation with a soft gluon emitted at large
    angle and a soft-collinear emitted gluon.
    That is precisely the method used in
    Ref.~\cite{FerrarioRavasio:2023kyg}.  }

Note that a formula essentially identical to Eq.~(\ref{eq:K-main}) has
appeared in Refs.~\cite{Hartgring:2013jma,Li:2016yez,Campbell:2021svd}
in the context of embedding NLO $Z \to 3\,\text{jets}$ within a shower
that was already NLO accurate for $Z \to 2\,\text{jets}$.
The presence of the same equation in both matching and shower kernels
is significant because it implies that the infrared limit of $K(z)$ in
matching will be by construction identical to the shower-kernel's
$K(z)$ (so long as the treatment of real radiation is also the same
between matching and shower).
This property is precisely what was identified in
Ref.~\cite{Hamilton:2023dwb} as being crucial in order to maintain
logarithmic accuracy when matching to fixed-order predictions and it
suggests that Eq.~(\ref{eq:K-main}) and its counterparts in
Refs.~\cite{Hartgring:2013jma,Li:2016yez,Campbell:2021svd} will play a
major role in generalising logarithmically accurate matching beyond
NLO.

\section{Evaluations of $K(z)$}
\label{sec:K-determination}

\subsection{$C^2_F$ term from a slicing scheme}
\label{sec:K-CF-slicing}

One general way of evaluating Eq.~(\ref{eq:K-main}) is to apply a
slicing scheme to the real integrals, splitting $K$ into two parts
\begin{align}
  \label{eq:K-two-parts}
  K(z) = K_<(z) + K_>(z)\,.
\end{align}
The first term contains the virtual corrections and the
unresolved real contributions and needs to be evaluated using
dimensional regularisation
\begin{subequations}
  \label{eq:K-sliced}
  \begin{align}
    \label{eq:Klt}
    K_<(z) &= \frac{V_{q\bar q g}}{B_{q\bar q g}}
               - \frac{V_{q\bar q}}{B_{q\bar q}}
               + \int^{\tilde \lambda}_0
               \frac{d\Phi_{q\bar q ij}}{d\Phi_{q\bar q g}}
               \frac{B_{q\bar q ij}}{B_{q \bar q g}}
               - \int^{\lambda}_{0}
               \frac{d\Phi_{q\bar q g'}}{d\Phi_{q\bar q}}
               \frac{B_{q\bar q g'}}{B_{q \bar q}}\,.
  \end{align}
The second term is purely finite and reads
  \begin{align}
    \label{eq:Kgt}
    K_>(z) &= 
    \int^{\tilde v_g}_{\tilde \lambda}
    \frac{d\Phi_{q\bar q ij}}{d\Phi_{q\bar q g}}
    \frac{B_{q\bar q ij}}{B_{q \bar q g}}
    - \int^{v_g}_{\lambda}
    \frac{d\Phi_{q\bar q g'}}{d\Phi_{q\bar q}}
    \frac{B_{q\bar q g'}}{B_{q \bar q}}\,.
  \end{align}
\end{subequations}
The $\tilde \lambda$ and $\lambda$ quantities represent the separation
scales in the phase spaces of the two distinct real terms.
As with the $\tilde v_g$ and $v_g$ constraints, they need to act
equivalently in the limit where $ij = g_1g_2$ and
$\theta_{g_2 q} \gg \theta_{g_1 q}$, in order to separately maintain
the process independence of both $K_<(z)$ and $K_>(z)$.
The physical scale associated with $\lambda$ should be much smaller
than the physical scale associated with $v_g$ to ensure that
$\lambda/v_g$ suppressed power corrections are small.

In practice, for the abelian $C^2_F$ term, denoted by $K^{(\rm ab)}$, we only need to consider the
$ij = g_1g_2$ channel.\footnote{
There is an additional $C_F^2$ contribution associated with the identical flavour $q\to q\bar{q}q$ splitting which is proportional to $C_F(2C_F - C_A)$. Our final results will be presented for $C_F=C_A/2$, eliminating this contribution.}
We take $g_2$ to be the lower-$v$ gluon and the
conditions we apply in Eqs.~(\ref{eq:Klt}) and (\ref{eq:Kgt}) are
\begin{subequations}
  \label{eq:lambda-conditions}
  \begin{align}
  \min(E_{g_2}, E_q) \theta_{g_2 q} &\lessgtr \tilde \lambda\,,
  \\
  \label{eq:lambda-conditions-b}
  \min(E_{g'}, E_q)  \theta_{g' q}  &\lessgtr \lambda\,,
  \end{align}
\end{subequations}
in analogy with the choice of ordering variable for our toy shower.
One should keep in mind that the energy of the quark in the two
conditions will in general be different.
In the soft limit for $g_2$ and $g'$, the conditions are, however,
independent of the quark energy, and we obtain the process
independence separately of both $K^{(\text{ab})}_<(z)$ and $K^{(\text{ab})}_>(z)$ by choosing
$\tilde \lambda = \lambda$.%
\footnote{If we had chosen invariant-mass conditions such as
  $E_{g_2} E_q \theta_{g_2 q} \lessgtr \tilde \lambda$ and analogously
  for the second condition, then we would lose the process
  independence unless we applied a suitable relative rescaling of the
  $\tilde \lambda$ and $\lambda$ to take into account the different
  quark energies. }

The $K_<^{(\text{ab})}(z)$ term is straightforward to evaluate analytically in
dimensional regularisation.
The details are provided in Appendix~\ref{sec:kl-eval}, with the slicing conditions of
Eq.~(\ref{eq:lambda-conditions}) and $\tilde \lambda$ set equal to
$\lambda$.
The result is
\begin{equation}
  \label{eq:Klt-result}
  K_<^{(\text{ab})}(z) =
  2 C_F \left[\text{Li}_2\left(-\frac{1-z}{z}\right)
    -\ln (z) \left(
      2 \ln \frac{v_g}{\lambda} +\ln
      \frac{z(1-z)}{\min (1-z,z)^2}
    \right)
  \right]
  -\frac{C_F}{p_{qq}(z)}\,.
\end{equation}

For the evaluation of $K_>^{(\text{ab})}(z)$, we simply re-use the code developed
for the toy shower as described in Section~\ref{sec:form-toy-show},
adapting it to carry out a single emission without a Sudakov form
factor, i.e.\ operating it as a fixed-order code.
In addition to the $\lambda$ cutoff, we would also need to place a
common cut on the maximum angle of any of $i,j$ and $g'$ so as to
ensure that we remain in the collinear regime.

\begin{figure}
  \centering
  \includegraphics[width=0.7\textwidth,page=1]{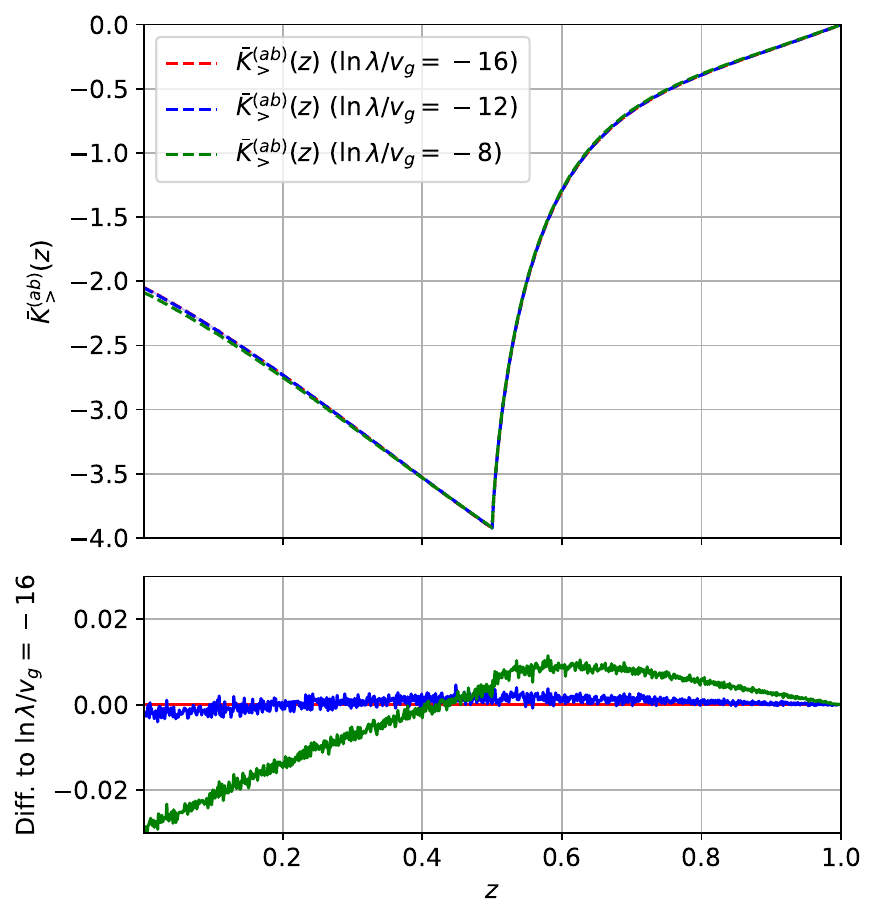}
  \caption{ %
    Top panel: Result for the $\bar{K}^{\rm(ab)}_>$ slicing
    contribution, for three different choices of
    $\ln\lambda / v_g\in [-8,-12,-16]$. We set $C_F = 4/3$ for the
    numerical evaluations.  Lower panel: Difference of
    $\bar{K}^{\rm(ab)}_>$ evaluated with the above values of the
    cutoff to the result with $\ln\lambda / v_g = -16$.
  }
  \label{fig:slicing-results}
\end{figure}

In practice, we find it is convenient to directly integrate the
difference of the integrands, rather than to evaluate the two
integrals separately.
The cut on $\lambda$ induces a term proportional to $\ln z \ln
v_g/\lambda$ in $K_>^{(\text{ab})}(z)$, which cancels against the
corresponding term in $K_<^{(\text{ab})}(z)$.
That term dominates in the statistical errors for
$K_>^{(\text{ab})}(z)$, as in any slicing approach.
To have better numerical behaviour, we instead directly
evaluate a quantity
\begin{equation}
  \bar{K}_>^{(\text{ab})}(z)
  = K_>^{(\text{ab})}(z) - 4 C_F \ln z \ln\frac{v_g}{\lambda}\,,
\end{equation}
using different choices of the large-angle cut
in the $1\to2$ and $2\to3$ phase spaces, so as to
subtract the logarithmic term.
Specifically we match the logarithmic phase space volume between the two terms of
Eq.~(\ref{eq:Kgt}) in the limit of strong $v$ ordering.
The results for $\bar{K}_>^{(\text{ab})}(z)$ are shown in
Fig.~\ref{fig:slicing-results}.
We show several values of $\lambda$ and, as expected, the dependence
on $\lambda$ becomes negligible as it is taken towards zero.
We defer discussion of the structure of $K^{(\text{ab})}(z)$ to
Section~\ref{sec:Kz-comments}.

\subsection{Relation of $K(z)$ with ${\cal B}^q_2(z)$ calculation }
In this section, we will show how the quantity $K(z)$ can be simply
related to the ${\cal B}_2(z)$ anomalous dimensions computed in
Refs.~\cite{Dasgupta:2021hbh,vanBeekveld:2023lsa}. In particular, for
the NS flavour channel we will use the quark ${\cal B}_2^q(z)$
calculation of Ref.~\cite{Dasgupta:2021hbh}.
For a quark, in the notation used for Eq.~\eqref{eq:K-main} this
quantity is defined as
\begin{equation}
  \label{eq:B-main}
  \frac{{\cal B}_{2,v}^q (z)}{P_{qq}(z)} =
  \frac{V_{q\bar q g}}{B_{q\bar q g}}
  - \frac{V_{q\bar q}}{B_{q\bar q}}
  + \int^{\tilde v_g}_0
  \frac{d\Phi_{q\bar q ij}}{d\Phi_{q\bar q g}}
  \frac{B_{q\bar q ij}}{B_{q \bar q g}}
  - \int^{\tilde v_g}_{0}\left[  \frac{d\Phi_{q\bar q ij}}{d\Phi_{q\bar q g}}
  \frac{B_{q\bar q ij}}{B_{q \bar q g}}
\right]_{\rm s.o.}\,.
\end{equation}
The previous equation is a generalisation of the definition used in
Refs.~\cite{Dasgupta:2021hbh,vanBeekveld:2023lsa} to a generic
ordering variable $v$, like that defined in
Eq.~\eqref{eq:vi-definition} (hence the $v$ subscript).
We see that the first three terms in Eqs.~\eqref{eq:K-main}
and~\eqref{eq:B-main} are in common, and the only difference, aside
from normalisation, stems from the last term in the r.h.s.\ of
Eq.~\eqref{eq:B-main}.
In Eq.~\eqref{eq:B-main}, the term labelled by
$\left[\dots\right]_{\rm s.o.}$ has the role of subtracting the
strongly angular ordered approximation of the $1\to 3$ splitting
kernel entering the third term in the r.h.s.\ of the equation.
In defining ${\cal B}_{2,v}^q (z)$ this is important so as to isolate NSL
contributions arising solely from the commensurate-angles region of
the $1\to 3$ phase space.

We will start with the discussion of the $C_F^2$ colour channel and
compare the result to what was obtained with the slicing scheme in the
previous section. We will then discuss the $C_F C_A$, $C_F T_R n_f$,
and $C_F(C_F-C_A/2)$ contributions (the latter will be eventually
neglected by taking the large-$N_c$ limit).
In the discussion below, it is instructive to study first the simpler
case of angular ordering, as originally considered in
Refs.~\cite{Dasgupta:2021hbh,vanBeekveld:2023lsa}. As we will show,
for angular ($\theta$) ordering the inclusive emission probability can
be directly taken from these references, since the difference between
$K(z)$ and ${\cal B}^q_2(z) \equiv {\cal B}_{2,\theta}^q (z)$
vanishes.
We then move on to discuss the case of the ordering variable in
Eq.~\eqref{eq:vi-definition}, used to obtain the numerical results of
this article. In this case, additional considerations w.r.t.\ the
angular ordering case are necessary, as will be discussed in the
present section. As we will show, the quantity $K(z)$ is related to
${\cal B}_{2,v}^q (z)$ via a finite correction term involving only
$1\to 2$ splitting kernels, which can be computed in four space-time
dimensions.

\subsubsection{Abelian $C_F^2$ contributions}
\label{sec:Kz-CF-analytical-scheme}
We first connect $K(z)$ defined in Eq.~\eqref{eq:K-main} and
${\cal B}_2^q(z)$ in the case of angular ordering, as initially
considered in Refs.~\cite{Dasgupta:2021hbh,vanBeekveld:2023lsa}.
To avoid confusion, we will use $K_{\theta}(z)$ to denote $K(z)$ in
Eq.~\eqref{eq:K-main} obtained with angular ordering, while the
notation $K(z)$ will be used for our default transverse-momentum
ordering case.

\paragraph{Angular ordering case:}
\newcommand{\ab}{C_F^2}
The definition of $K_{\theta}(z)$ in Eq.~\eqref{eq:K-main} can be
easily related to ${\cal B}_{2,\theta}^q(z)$ in the $C_F^2$ channel
(${\cal B}_{2,\theta}^{q, \ab}(z)$), defined for instance in
Eq.~(3.45) of Ref.~\cite{Dasgupta:2021hbh} and reported in
Appendix~\ref{sec:B2} (we remind the reader that the identical-quark
splitting $q\to q\bar{q} q$, which would normally give a
$C_F (C_F-C_A/2)$ contribution, is zero in the leading-$N_c$
approximation that we use here).
Aside from the overall normalisation factor $P_{qq}(z)$ mentioned
above, the only difference between the two quantities is encoded in
the last term of Eq.~\eqref{eq:K-main} vs.\ the subtraction of the
strongly-ordered contribution to ${\cal B}_{2,\theta}^q(z)$ in
Eq.~\eqref{eq:B-main} (see also Eqs.~(3.43) and~(3.45) of
Ref.~\cite{Dasgupta:2021hbh}).
As stressed above, this term has the role of subtracting SL physics of
strongly-ordered origin.
Specifically, a first difference is that, in the calculation of
${\cal B}_{2,\theta}^{q,\ab}(z)$,
the strongly-ordered subtraction is defined by
integrating the product of leading-order splitting kernels with an
ordering condition that coincides with the physical ordering, that is
$\theta_{g_p q_i} > \theta_{g_i q_i}$. This corresponds to the
$\tilde{v}_g$ physical ordering mentioned in
Section~\ref{sec:virtual-general}.
Conversely, the calculation of the strongly-ordered subtraction in
$K^{(\text{ab})}_{\theta}(z)$ is performed with the shower ordering $v_g$, that in
our case is equivalent to the condition
$\theta_{g_p q_q} > \theta_{g_i q_i}$
, where $\theta_{g_p q_p}$ is
the angle between the first gluon and the intermediate quark that
splits into $g_i q_i$.
Secondly, while the strongly-ordered subtraction in the calculation of
${\cal B}_{2,\theta}^{q,\ab}(z)$ adopts the matrix element with collinear
singularities given by $1/\theta_{g_p q_i}^2\times 1/\theta_{g_i q_i}^2$,
the same ingredient in Eq.~\eqref{eq:K-main} uses the shower matrix
element with singularities
$1/\theta_{g_pq_p}^2\times 1/\theta_{g_i q_i}^2$.

It is then clear that the difference between
${\cal B}_{2,\theta}^{q,\ab}(z)$ and $K_{\theta}^{(\text{ab})}(z)$ is a term of
strongly-ordered origin that is finite and can be calculated in four dimensions.
We use the notation
\begin{align}~\label{eq:DB2}
\Delta {\cal B}_{2,\theta}^{q,\ab} \equiv \frac{P_{qq}(z)}{C_F^2} K_{\theta}^{\rm
  (ab)}(z)-{\cal B}_{2,\theta}^{q,C_F^2}(z) \,,
\end{align}
to parameterise the difference from $K_{\theta}^{\rm(ab)}(z)$ in the abelian colour
channel.
For angular ordering, an explicit calculation leads to
\begin{align}\label{eq:DB2-theta}
  \Delta {\cal B}_{2,\theta}^{q,\ab} = p_{qq}(z)\left(\int
  \frac{d\theta_{g_p q_i}^2}{\theta_{g_p q_i}^2} \frac{d\theta_{g_i
  q_i}^2}{\theta_{g_i q_i}^2}\,d z_i\,
  \frac{d  \Delta \phi}{2\pi}\,p_{qq}(z_i)\theta^2\delta(\theta^2-\theta^2_{g_p
  q_i})\Theta(\theta_{g_p q_i} - \theta_{g_i q_i}) \right.\notag\\
  \left. - \int
  \frac{d\theta_{g_p q_p}^2}{\theta_{g_p q_p}^2} \frac{d\theta_{g_i
  q_i}^2}{\theta_{g_i q_i}^2}\,d z_i\,\frac{d \Delta \phi}{2\pi}\,p_{qq}(z_i)\theta^2\delta(\theta^2-\theta^2_{g_p q_p})\Theta(\theta_{g_p q_p} - \theta_{g_i q_i}) 
\right)=0\,.
\end{align}
Here $\theta$ denotes the angle that is kept fixed in the definition
of the Sudakov integrand, which corresponds to the shower evolution
variable. In the first (second) integral, $\Delta \phi$ is the azimuthal angle between the plane containing $g_p$ and $q_i$ ($q_p$), and that containing $g_i$ and $q_i$.
Eq.~\eqref{eq:DB2-theta} amounts to stating that, as demonstrated in
Refs.~\cite{Dasgupta:2021hbh,vanBeekveld:2023lsa}, for an
angular-ordered shower the quantity ${\cal B}_{2,\theta}^{q,\ab}(z)$
coincides (up to a normalisation) with $K^{\rm(ab)}_\theta(z)$ needed for NSL
accuracy.
The explicit expression for $ {\cal B}_{2,\theta}^{q,\ab} (z)$ is
given in Appendix~\ref{sec:B2}.

\paragraph{Transverse-momentum ordering case:}
We now switch to the case where the emissions are ordered in the
transverse momentum variable defined in
Eq.~\eqref{eq:vi-definition}. The quantity ${\cal B}_2^q(z)$ for this
ordering variable in the abelian channel,
${\cal B}_{2,v}^{q, \ab}(z)$, can be expressed as
(cf. Appendix~\ref{sec:B2})
\begin{equation}\label{eq:B2def-v}
{\mathcal B}^{q, \ab}_{2,v}(z) = {\mathcal B}^{q,\,{\rm an.},\ab}_{2}(z) + C_F^2  H_{v}^{\rm fin.}(z)\,.
\end{equation}
The term $ {\mathcal B}^{q,\,{\rm an.},\ab}_{2}(z) $ is the same as
for angular ordering, first computed in Ref.~\cite{Dasgupta:2021hbh}, and the result is reported in
Eqs.~\ref{eq:B2qCFCF}.

The quantity $H_{v}^{\rm fin.}(z)$ can instead be simply calculated numerically. Here
we follow the same procedure used for the angular ordering case in
Refs.~\cite{Dasgupta:2021hbh,vanBeekveld:2023lsa}, discussed in
Appendix~\ref{sec:B2}.
One key property of ${\cal B}_2^q(z)$ is that it describes the physics
of emissions at commensurate angles. In the explicit calculation, this
requires the careful removal, from the triple-collinear splitting
function, of the disparate angle configurations associated to
single-logarithmic terms. %
In the transverse momentum ordering case, this subtraction involves removing
both the sectors $\theta_{g_p q_i} \gg \theta_{g_i q_i}$ and
$\theta_{g_p q_i} \ll \theta_{g_i q_i}$, which are allowed by the
transverse momentum ordering condition $v_{g_p q_i} > v_{g_i q_i}$.
This leads to a result for $H_{v}^{\rm fin.}(z)$ that reads \footnote{For the angular ordering case, the analogous result is given in  Eq.~\eqref{eq:Hfin-theta}, where the
second term in brackets removes the configurations where
$\theta_{g_p q_i} \gg \theta_{g_i q_i}$.}
\begin{align}\label{eq:Hfin-v}
  H_{v}^{\rm fin.}(z) &=  \int
  d\Phi_{3}(8 \pi \alpha_s)^2 \Theta(v_{g_p q_i}
    - v_{g_i q_i})\,\delta(z - z_p)\,v\,\delta(v-v_{g_pq_i})\\&\times\bigg(\frac{\langle
  P\rangle_{C_F^2}}{s_{g_p g_i q_i}^2} -  \Theta(\theta_{g_p q_i} - \theta_{g_i q_i})\left[\frac{\langle P\rangle_{C_F^2}}{s_{g_p g_i
        q_i}^2}\right]_{ \theta_{g_p q_i}\gg \theta_{g_i q_i}} \hspace{-1cm}- \Theta(\theta_{g_i q_i} - \theta_{g_p q_i})\left[\frac{\langle P\rangle_{C_F^2}}{s_{g_p g_i
        q_i}^2}\right]_{\theta_{g_p q_i}\ll \theta_{g_i q_i}} \bigg)\,.\notag
\end{align}
The three-body phase space $d\Phi_{3}$ is explicitly defined in
Eq.~\eqref{eq:psB} and the spin averaged $1\to 3$ splitting function
$\langle P\rangle_{C_F^2}=\langle
P\rangle_{C_F^2}(x_{g_p},x_{g_i},x_{q_i})$ (with $x_i$ being the
longitudinal momentum fraction of particle $i$ w.r.t.~the energy of
the initiating parton) can be found in Ref.~\cite{Catani:1998nv}.

\begin{figure}[h!]
  \centering
  \includegraphics[width=0.48\textwidth]{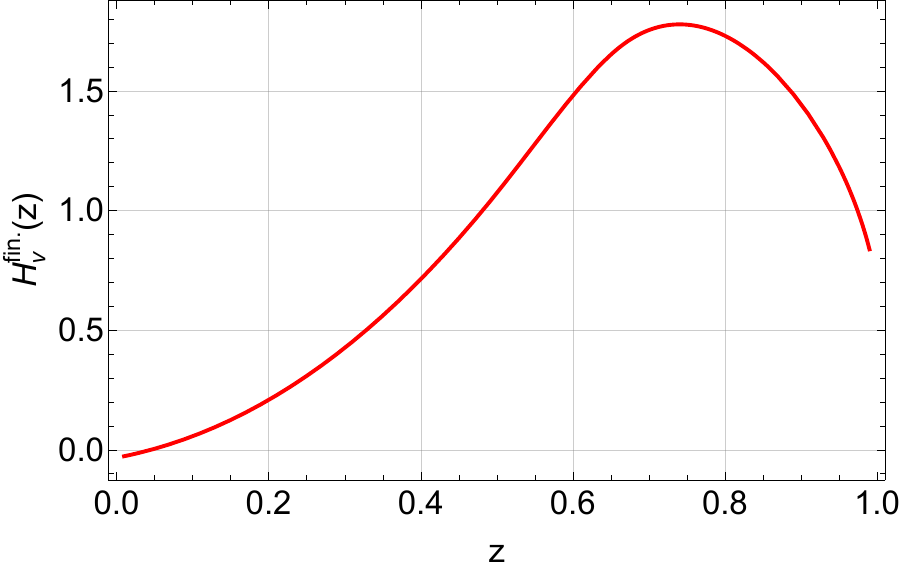}
  \includegraphics[width=0.48\textwidth]{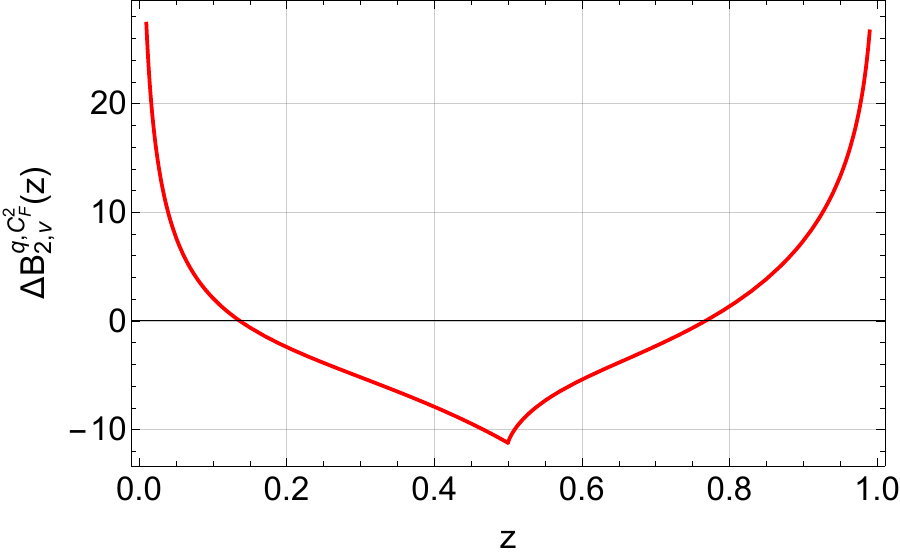}
  \caption{The functions $H_{v}^{\rm fin.}(z)$ and $\Delta {\cal B}_{2,v}^{q,\ab} $.}
  \label{fig:Hfin-kt}
\end{figure}

The two strongly-ordered counter-terms in the last line amount to
subtracting the leading term in the expansion of  $\frac{\langle
  P\rangle_{C_F^2}}{s_{g_p g_i q_i}^2} $ in the limits $\theta_{g_p
  q_i}\gg \theta_{g_i q_i}$ and $\theta_{g_p q_i}\ll \theta_{g_i q_i}$.
This is crucial for $ H_{v}^{\rm fin.}(z) $ to describe collinear
dynamics at commensurate angles. Given our phase-space parametrisation 
(see also Eq.~\eqref{eq:mappingxtoz})
\begin{equation}\label{eq:B-splits}
x_{g_p} = 1-z_p\,,\quad x_{g_i} = z_p \,(1-z_i),\quad x_{q_i}=z_p\,z_i\,,
\end{equation}
we obtain the following explicit expressions:
\begin{align}\label{eq:Pcf2}
\left[\frac{\langle P\rangle_{C_F^2}}{s_{g_p g_i
        q_i}^2}\right]_{\theta_{g_p q_i}\gg \theta_{g_i q_i}} &= \frac{{\cal
  J}_1(z_p,z_i)}{E^4}\,\frac{P^{(0)}_{qq}(1-x_{g_p})}{\theta_{g_p q_i}^2}
\frac{P^{(0)}_{qq}(1-x_{g_i}/(1-x_{g_p}))}{\theta_{g_i q_i}^2}\,,\\
\left[\frac{\langle P\rangle_{C_F^2}}{s_{g_p g_i
        q_i}^2}\right]_{\theta_{g_p q_i}\ll \theta_{g_i q_i}} &= \frac{{\cal
  J}_2(z_p,z_i)}{E^4}\, \frac{P^{(0)}_{qq}(1-x_{g_i})}{\theta_{g_p q_i}^2}
\frac{P^{(0)}_{qq}(1-x_{g_p}/(1-x_{g_i}))}{\theta_{g_i q_i}^2}\,,
\end{align}
where we defined
\begin{align}\label{eq:JPS}
  {\cal J}_1(z_p,z_i)
  &\equiv
    \frac{1}{(1-z_p) z_p^3 (1-z_i) z_i },\\
  {\cal J}_2(z_p,z_i)
  &\equiv
    \frac{z_p}{1-z_p(1-z_i)}{\cal J}_1(z_p,z_i)%
  \,.
\end{align}
We calculate Eq.~\eqref{eq:Hfin-v} numerically, and its result is
shown in Figure~\ref{fig:Hfin-kt} (left).

As for the angular ordering case, the difference in the $C_F^2$
channel between ${\cal B}_{2,v}^{q,\,\ab}(z)$ and $K^{\rm (ab)}(z)$
amounts to an integral over the iterated $1\to 2$ splitting function,
while the genuine $1\to 3$ contribution cancels between the two
quantities. We then define
\begin{align}~\label{eq:DB2-v}
\Delta {\cal B}_{2,v}^{q,\ab} \equiv \frac{P_{qq}(z)}{C_F^2} K^{\rm
  (ab)}(z)-{\cal B}_{2,v}^{q,\ab}(z) \,.
\end{align}
The difference $\Delta {\cal B}_{2,v}^{q,\ab} $ stems from the
different definition of the last term in the r.h.s. of
Eqs.~\eqref{eq:K-main},~\eqref{eq:B-main}, and we calculate it
analytically. Its expression is reported as an ancillary file with
this article and it is shown in Fig.~\ref{fig:Hfin-kt} (right).
\begin{figure}[h!]
  \centering
  \includegraphics[page=3,width=0.48\textwidth]{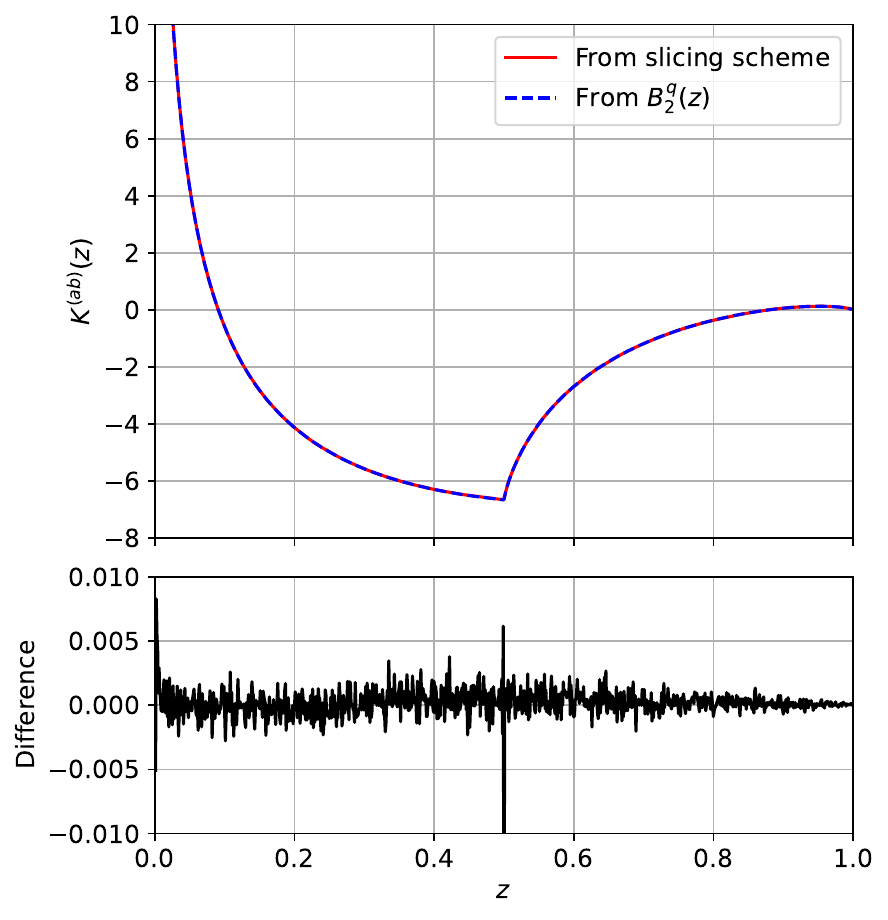}
  \caption{$K^{\rm{(ab)}}(z)$ calculated both from the slicing scheme of section~\ref{sec:K-CF-slicing} and from ${\cal B}_{2,v}^{q,C_F^2}(z)$. }
  \label{fig:K_from_B2}
\end{figure}
The plot shown in Fig.~\ref{fig:K_from_B2} demonstrates that the two
calculations of $K^{\rm{(ab)}}(z)$ are in agreement at the per mil level.

Note that for the algorithm considered in this article, the shower-dependent
quantity $\Delta {\cal B}_{2,v}^{q,\ab}$ depends only on
strongly-ordered $1\to 2$ splitting kernels, making it very simple to
calculate. This may not be the case in general for other flavour
channels, depending on the details of the implementation of the full
$1\to 3$ splitting in the shower.

\subsubsection{Remaining colour channels}
\label{sec:Kz-CAnf-terms}
We now discuss the remaining channels contributing to
$K(z)$~\eqref{eq:K-main}, including contributions from $C_A$,
$T_R n_f$ and $C_F-C_A/2$ colour structures. A first aspect to
consider is that $K(z)$ agrees with $K_{\rm CMW}$~\cite{Catani:1990rr} in the soft limit,
that is
\begin{equation}
K(z) = K_{\rm CMW}+\tilde{K}(z)\,,
\end{equation}
where $\tilde{K}(z)$ vanishes in the soft limit and it is integrable
in $z\in [0,1]$.
A second consideration is that, in the non-singlet channel considered
here with $C_F = C_A/2$, the only flavour structures that modify the
quark momentum are those related to consecutive independent (abelian)
splittings, considered in the previous section. Therefore, in the NS
channel, the contribution to $\tilde{K}(z)$ from splittings
$q\to q\, p_1\, p_2$ not contributing to the abelian $C_F^2$ channel,
can be obtained by fixing the kinematics of the parent $p_1+p_2$ and
integrating inclusively over the remaining phase space. This is
precisely the definition of ${\cal B}_{2,v}^q$ given in
Appendix~\ref{sec:B2}. This gives a straightforward relationship
between $\tilde{K}(z)$ and ${\cal B}_{2,v}^q$, which coincide, up to
the usual $P_{qq}(z)$ normalisation, in the $C_A$, $T_R n_f$ and
$C_F-C_A/2$ colour channels.
This simple relationship will be modified once the
  non-abelian colour channels are included in the shower's $1\to 3$
  real radiation, as was the case for the $C_F^2$
  contribution discussed in detail in the previous subsection. The methods presented in this article readily
  apply to those cases as well. 
  Specific classes of observables, such as event shapes, are by
  construction not sensitive to the $z$ dependence of
  ${\cal B}_{2,v}(z)$, but only to its inclusive integral. In these
  cases, a simplified procedure to relate the inclusive emission
  probability in the shower to the integral of ${\cal B}_{2,v}(z)$
  was presented in Ref.~\cite{vanBeekveld:2024wws} in the context of
  NNLL-accurate parton showers.

\subsection{Comments on structure of $K(z)$}
\label{sec:Kz-comments}

Here we comment on the structure of the $C_F^2$ contribution to
$K(z)$, as shown in Fig.~\ref{fig:K_from_B2}.
Firstly, we observe that $K(z)$ goes to zero for $z \to 1$.
This is as expected, since at large $z$, the full $K(z)$ should
reduce to the well known $\Kcmw$ term~\cite{Catani:1990rr},
which has $C_A$ and $T_R n_f$ contributions, but no $C_F$
contribution.

Secondly, it is continuous but not smooth at $z = 1/2$.
In the slicing version of the calculation, such structure is
explicitly seen in Eq.~(\ref{eq:Klt-result}), and is present also in
$K_>(z)$. It connects with the use of a renormalisation scale
choice $\mu = E\min(z,1-z)\theta_{qg}$ and with the ordering
condition in Eqs.~(\ref{eq:real-1to3}) and (\ref{eq:vi-definition}),
both of which involve non-smooth behaviour at $z=1/2$.

Next, we note that there is a $-C_F \ln^2 (z)$ behaviour for
$z \to 0$, which in the slicing calculations originates from
$K^{(\text{ab})}_<(z)$.
This small-$z$ integrable divergence was also already identified in
${\cal B}_{2,\theta}^{q,\ab}(z)$ in Ref.~\cite{Dasgupta:2021hbh}.  The
structure for $z \to 0$ is connected with the soft-quark limit and
while interesting in its own right, goes beyond the scope of this
article.

A crucial property of ${\cal B}_{2,v}(z)$ is that, analogously to its
angular-ordered
counterpart~\cite{Dasgupta:2021hbh,vanBeekveld:2023lsa}, it satisfies
the sum rule
\begin{equation}\label{eq:sum-rule}
  \int_0^1 d z \,{\cal B}^q_{2,v}(z) = -\gamma_q + b_0 X_v\,,
\end{equation}
where $\gamma_q$ is the endpoint of time-like DGLAP anomalous
dimension and the constant $X_v$ depends on the chosen ordering
variable, which can be easily obtained by integrating the expressions
reported in Appendix~\ref{sec:B2}. In the abelian channel this implies
that ${\cal B}^q_{2,v}(z) $ should integrate to the corresponding
$C_F^2$ term of $-\gamma_q$. This sum rule should not be modified by
the contributions relating ${\cal B}^q_{2,v}(z)$ with $K(z)$, which we
have verified is the case.

This is consistent with the expectations of
Ref.~\cite{vanBeekveld:2024wws}, specifically Eqs.~(3), (6) and (29).
To help see this, it is useful to consider $\Delta_{\ln z}$ in Eq.~(6)
of that reference (with $z$ there corresponding to $1-z$ in this
work).
This quantity is the coefficient of a drift in $\ln(1-z)$ (using our
notation) of a soft-collinear emission as induced by a subsequent
splitting.
In the soft-collinear limit, our shower leaves the kinematics of the
soft-collinear gluon unchanged and therefore the drift is identically
zero, and this results in the analogue of Eq.~(\ref{eq:sum-rule}) for
$K(z)$.
Note that the reason why one trivially has a zero drift here is that
our real emission map preserves the kinematics of a first
soft-collinear emission, while this is not the case for the full
showers considered in Ref.~\cite{vanBeekveld:2024wws}.
%

\section{Logarithmic tests}
\label{sec:log-tests}

Our implementation of the collinear shower does not include matching
to NLO $2$-jet production, which effectively corresponds to a
coefficient function at the high scale.
To test the shower, rather than starting the evolution from that high scale and
including the corresponding coefficient function, we will consider
the evolution between two disparate infrared scales and compare that
result to a semi-analytic reference calculation obtained using a
specifically adapted version of the HOPPET DGLAP evolution
code~\cite{Salam:2008qg}, using the ingredients from
Refs.~\cite{Curci:1980uw,vanBeekveld:2024jnx}.
We will carry out two sets of non-singlet tests, one on the
fragmentation function, the other on the spectrum of small-$R$ jets.

\subsection{Non-singlet fragmentation function}
\label{sec:result-non-singlet-frag}

We start by considering the non-singlet fragmentation function. In the
collinear shower, this quantity is computed as follows. We start the
shower evolution at a resolution scale $v_{\rm max}$, where the final
state consists of a single quark of momentum fraction $z=1$. We then
evolve down to a second scale $v_{\rm min} \ll v_{\rm max}$, at which
we measure the $z$ distribution of the final quark.\footnote{The proper
  definition of the non-singlet fragmentation function would involve
  the difference in the $z$ distribution between quarks and
  anti-quarks. However, since our proof-of-concept collinear shower
  does not contain $g\to q\bar{q}$ splittings, there are no
  anti-quarks in the final state.}
We denote this quantity by $D^{\rm (PS)}_{\rm NS}(z,v_{\min},v_{\max})$.

To validate the shower result, we compare the resulting $z$
distribution to a reference NSL calculation. Results at NSL for the
fragmentation functions are known in the common $\MSbar$ scheme, where
the collinear singularities are regularised in $4-2\epsilon$
space-time dimensions. 
In order to connect to the shower result, we thus need to perform a
scheme change, that can be implemented via a matching
coefficient.
Taking this into account, the analytic expectation for $D^{\rm
  (PS)}_{\rm NS}(z,v_{\min},v_{\max})$ is given by
\begin{equation}
  \label{eq:DNS-reference}
  D^{\rm
    (NSL)}_{\rm NS}(z,v_{\min},v_{\max}) = C(v_{\min}) \otimes
  \exp\left [\int_{v_{\min}^2}^{v_{\max}^2} \frac{dv^2}{v^2} 
    \hat P(v)
  \right]
  \otimes C^{-1}(v_{\max})\,.
\end{equation}
In the above equation,
$\hat{P}(v) = \as(v) \hat{P}^{(0)}(z)/2\pi + \as^2(v)
\hat{P}^{(1,\MSbar)}(z)/4\pi^2$ denotes the standard DGLAP non-singlet anomalous
dimension in the $\MSbar$ scheme~\cite{Curci:1980uw} (the path
ordering can be dropped in the non-singlet channel).
The matching coefficient $C(v)$ encodes the scheme change
from $\MSbar$ to the shower regularisation scheme. At NSL, this is obtained
by computing the order $\as$ contribution to the NS fragmentation
function arising from the kinematical region below the shower cutoff.
For the evolution variable defined in Eq.~\eqref{eq:vi-definition},
the NSL matching coefficient reads
\begin{equation}
C(v) \equiv \delta(1-z) + \frac{\alpha_s(v)}{2\pi} C^{(1)}(z) + {\cal O}(\alpha_s^2)\,,
\end{equation}
where
\begin{multline}
C^{(1)}(z) =- 2\,P_{qq}(z)\left(\ln(1-z)\Theta\left(\frac12-z\right) +
             z\leftrightarrow (1-z)\right) - C_F\,(1-z)
  \,+ \\
           -\frac{C_F}{6}\,\left(2\,\pi^2-3\,\left(7-6\,\ln
   2\right)\right)\,\delta(1-z)\,.
\end{multline}
Eq.~\eqref{eq:DNS-reference} thus converts the initial condition
$\delta(1-z)$ from the shower scheme to $\MSbar$ at the higher scale
$v_{\max}$, it then evolves the resulting $\MSbar$ fragmentation function
down to $v_{\min}$, and finally it converts back to the shower scheme
at the end of the evolution.

When comparing Eq.~\eqref{eq:DNS-reference} to the shower result,
there will inevitably be subleading, NNSL, differences.
To single out the pure SL and NSL contributions, we use a standard technique~\cite{Dasgupta:2020fwr}  which relies on a small-$\as$ limit, while holding fixed the variable
\begin{equation}
  \label{eq:lambda-ff-def}
  \lambda=\as(v_{\max})\ln \frac{v_{\min}}{v_{\max}}\,.
\end{equation}
Specifically, we examine the quantity
\begin{equation}\label{eq:ratio-of-Ds}
 \frac{D^{\rm (PS)}_{\rm NS}(z,v_{\min},v_{\max})}{D^{\rm
      (NSL)}_{\rm NS}(z,v_{\min},v_{\max})} -1\,.
\end{equation}
To test the SL accuracy of the collinear shower we can take the
$\as\to 0$ limit of Eq.~\eqref{eq:ratio-of-Ds}.
Similarly, the NSL accuracy can be tested by dividing
Eq.~\eqref{eq:ratio-of-Ds} by $\as$ and then taking the $\as\to 0$
limit.
In both cases, we expect the result to be consistent with zero.

\begin{figure}[h!]
  \centering
  \includegraphics[page=2,width=\textwidth]{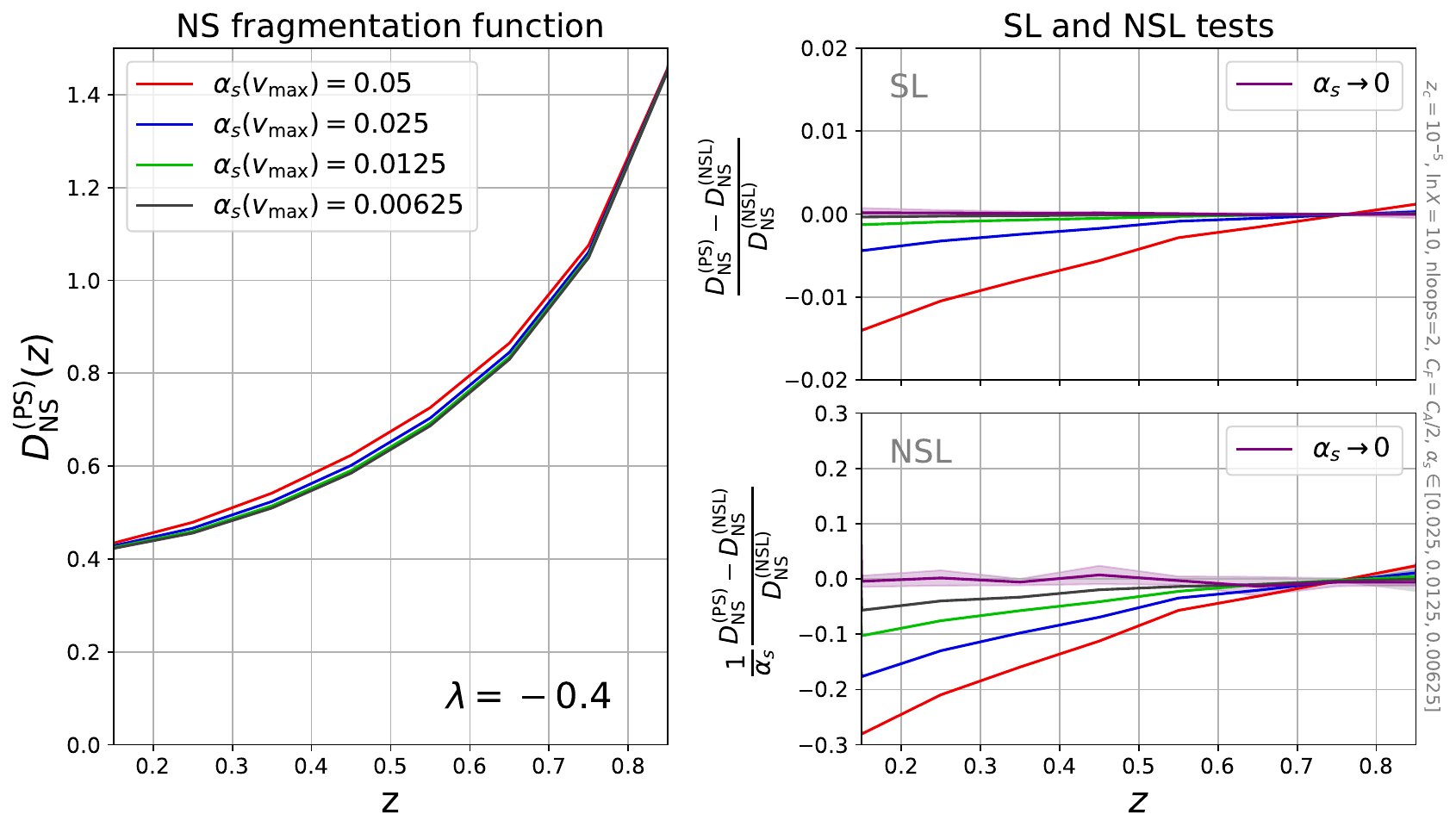}
  \caption{Shower non-singlet fragmentation function and associated
    logarithmic tests at $\lambda = -0.4$, cf.\ Eq.~(\ref{eq:lambda-ff-def}).
    The left-hand panel shows the non-singlet fragmentation function as obtained
    with the collinear shower, for four values of the coupling.
    The upper-right panel shows a test of the shower's SL accuracy,
    where the $\as \to 0$ extrapolation gives a result consistent with
    zero, as expected for SL accuracy.
    The lower-right panel shows a test of the shower's NSL accuracy,
    illustrating that the $\as \to 0$ extrapolation is consistent with
    zero, as required for NSL accuracy.
  }
  \label{fig:frag-func-test}
\end{figure}

The results of these SL and NSL accuracy tests are displayed in
Fig.~\ref{fig:frag-func-test}. As already explained in
Sec.~\ref{sec:form-toy-show}, here we adopt the large-$N_c$ limit in
which we set $C_F=C_A/2=3/2$. In this limit, the contribution from the $q\to qq\bar{q}$ splitting
(with identical flavours) vanishes.
The left plot in Fig.~\ref{fig:frag-func-test} shows the fragmentation
function $D^{\rm (PS)}_{\rm NS}(z,v_{\min},v_{\max})$ for a set of
small values of $\as\equiv\as(v_{\max})$.
The right plots show the ratio~Eq.\eqref{eq:ratio-of-Ds} (top panel) and
the same quantity divided by $\as$ (bottom panel) for each of the
above $\as$ values, and their extrapolation to $\as=0$.
In both cases, the extrapolated result agrees with zero to within
statistical and extrapolation uncertainties, therefore
showing consistency of the collinear shower with NSL accuracy.
Additional technical details on the setup used in this test are given
in Appendix~\ref{sec:further-technical-details}.

\begin{figure}
  \centering
  \includegraphics[page=2,width=0.5\textwidth]{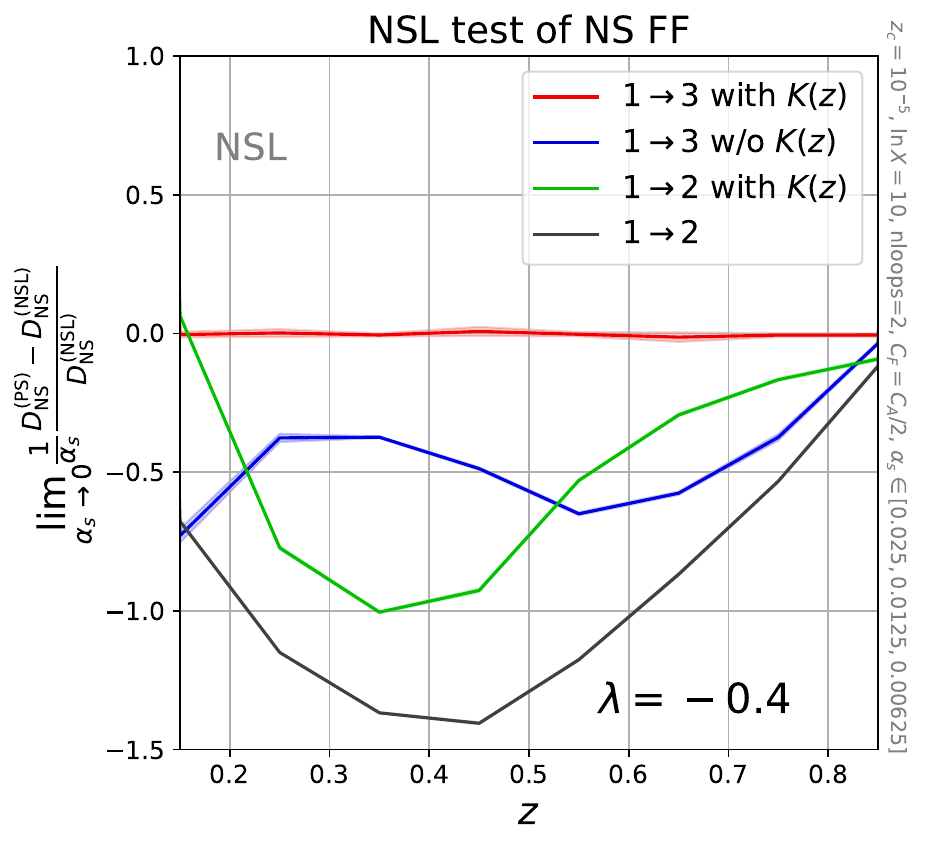}
  \caption{Illustration of the size of the NSL discrepancy in the
    non-singlet fragmentation function that arises when various NSL
    elements are left out.
    All curves already have the $\as \to 0$ limit taken.
  }
  \label{fig:fragfn-decomp}
\end{figure}

To help appreciate the separate numerical impact of the
triple-collinear and the inclusive NLO corrections,
Fig.~\ref{fig:fragfn-decomp} shows the size of the NSL discrepancy
that arises when various NSL contributions are turned off in the
shower.
The black curve corresponds to having just iterated $1\to2$ branchings
(strictly ordered in the shower evolution variable $v$) and no
inclusive $K(z)$ contribution.
It demonstrates that the missing NSL terms can be up to
$1.4\times\as$, which would suggest missing effects of order $10{-}20\%$ in
a phenomenological context.
The green curve shows the result when one maintains the $1\to2$
structure but includes $K(z)$, while the blue curve uses the iterated
$1\to 3$ splitting, but without $K(z)$.
These two curves show that the two sets of contributions
are comparable in magnitude.
The red curve includes all NSL contributions in the shower and is
consistent with zero (it is identical to the purple curve of the
lower-right panel of Fig.~\ref{fig:frag-func-test}).

\subsection{Non-singlet small-$R$ jets}
\label{sec:result-non-singlet-smallR}

Next, we consider an observable that is akin to the $z$ spectrum of
small-$R$ jets inside a fat jet of radius $R_0$ and energy $E$, with
$R\ll R_0 \ll 1$.
We take a quark of energy $E$ and start the shower evolution at a
sufficiently high scale $v_{\max}$ such that for any
$z\in [z_c,1-z_c]$ the angular scale $R_0$ can be generated by the
shower, where $z_c \ll 1$ is a cut-off on the splitting variable generated by the shower.
Any emissions at angles larger than $R_0$ are vetoed, so as to mimic
starting with a jet of radius $R_0$ and energy $E$, which can be
thought of as an initial fragmentation function equal to
$\delta(1-z)$.
We continue the evolution down to a scale $v_{\min}$ such that for any
$z\in [z_c,1-z_c]$, the small angular scale $R$ can be reached by the
shower.
This generates an ensemble of gluons plus one quark. We cluster that
set of particles using the Cambridge/Aachen
algorithm~\cite{Dokshitzer:1997in,Wobisch:1998wt} with jet radius $R$
as implemented in FastJet~\cite{Cacciari:2011ma}. We then identify the
jet that contains the quark, and we histogram its momentum fraction
$z$.\footnote{Were our shower to include also $g\to q\bar{q}$
  splittings, it would be necessary to use a jet algorithm that allows
  for an IRC-safe definition of jet flavour in the presence of
  soft-quark
  pairs~\cite{Banfi:2006hf,Banfi:2007gu,Czakon:2022wam,Gauld:2022lem,Caola:2023wpj},
  see also
  Refs.~\cite{Caletti:2022hnc,Caletti:2022glq,Larkoski:2024nub,Larkoski:2023upz}
  for related work.}
We refer to the resulting distribution as
$D_\text{R}^{(\text{PS})}(z,ER, E R_0)$.  Further technical details on
the construction of the observable from the results of the collinear
shower are given in Appendix~\ref{sec:smallR-angle-treatment}.

The reference analytic calculation for this observable can be derived
from existing work on microjet fragmentation
functions~\cite{Dasgupta:2014yra,Kang:2016mcy,vanBeekveld:2024jnx}. At
the NSL order, the result can be deduced from
Ref.~\cite{vanBeekveld:2024jnx}, 
giving
\begin{equation}
\label{eq:DR-reference}
D^{\rm
  (NSL)}_{\rm R}(z,ER,E R_0) = C^{(R)}(E R) \otimes
\exp\left [2\int_{ER}^{ER_0} \frac{d\mu}{\mu} 
  \hat P^{(R)}(\mu,ER)
\right]
\otimes [C^{(R)}(E R_0)]^{-1}\,,
\end{equation}
with
\begin{equation}
  \label{eq:PR}
  \hat{P}^{(R)}\!\left(\mu,E\,R\right) = \frac{\alpha_s(\mu^2)}{2\pi}
  \left(\hat{P}^{(0)}(z)+\frac{\alpha_s(\mu)}{2\pi}
    \hat{P}^{(1,\MSbar)}(z) -\frac{\alpha_s(ER)}{2\pi}
                                       \delta\hat{P}_{ik}^{(1)}\right)\,, \\
\end{equation}
and
\begin{subequations}
  \begin{align}\label{eq:deltaP}
    \delta \hat{P}_{qq}^{(1)} (z) &\equiv \left(2\,\ln z\, \hat{P}_{qq}^{(0)}\right)\otimes \hat{ P}_{qq}^{(0)}\\
                                  & = -C_F^2\,\ln z \left(\frac{3\,z^2+1}{1-z} \ln z - 4
                                    \frac{1+z^2}{1-z}\ln(1-z) - \frac{z (4+z)+1}{1-z}\right)\,.
  \end{align}
\end{subequations}
The coefficient function $C^{(R)}(\mu) = \delta(1-z) + \as(\mu)
C^{(1,R)}(z)/2\pi$ involves
\begin{multline}
  \label{eq:C1R}
  C^{(1,R)}(z) =
  -2\,C_F\,(1+z^2)\,\left(\frac{\ln(1-z)}{1-z}\right)_+ - C_F\,\left((1-z)+2\frac{1+z^2}{1-z}\,\ln
    z\right)\\
  +C_F\,\left(\frac{13}{2}-\frac{2}{3}\,\pi^2\right)\,\delta(1-z)\,.
\end{multline}
The main difference between Eq.~(\ref{eq:DR-reference}) and the
expressions of Ref.~\cite{vanBeekveld:2024jnx} is in the $[C^{(R)}(E
R_0)]^{-1}$ factor, which accounts for the starting point of the
shower, namely a jet of radius $R_0$ with energy $E$, as opposed to
the NLO hard matching coefficient for $e^+e^- \to q\bar q $ that appears
in Ref.~\cite{vanBeekveld:2024jnx}.
Recall that we consider only the non-singlet contributions, as in
Appendix~C of that reference.

\begin{figure}[h!]
	\centering
	\includegraphics[page=2,width=\textwidth]{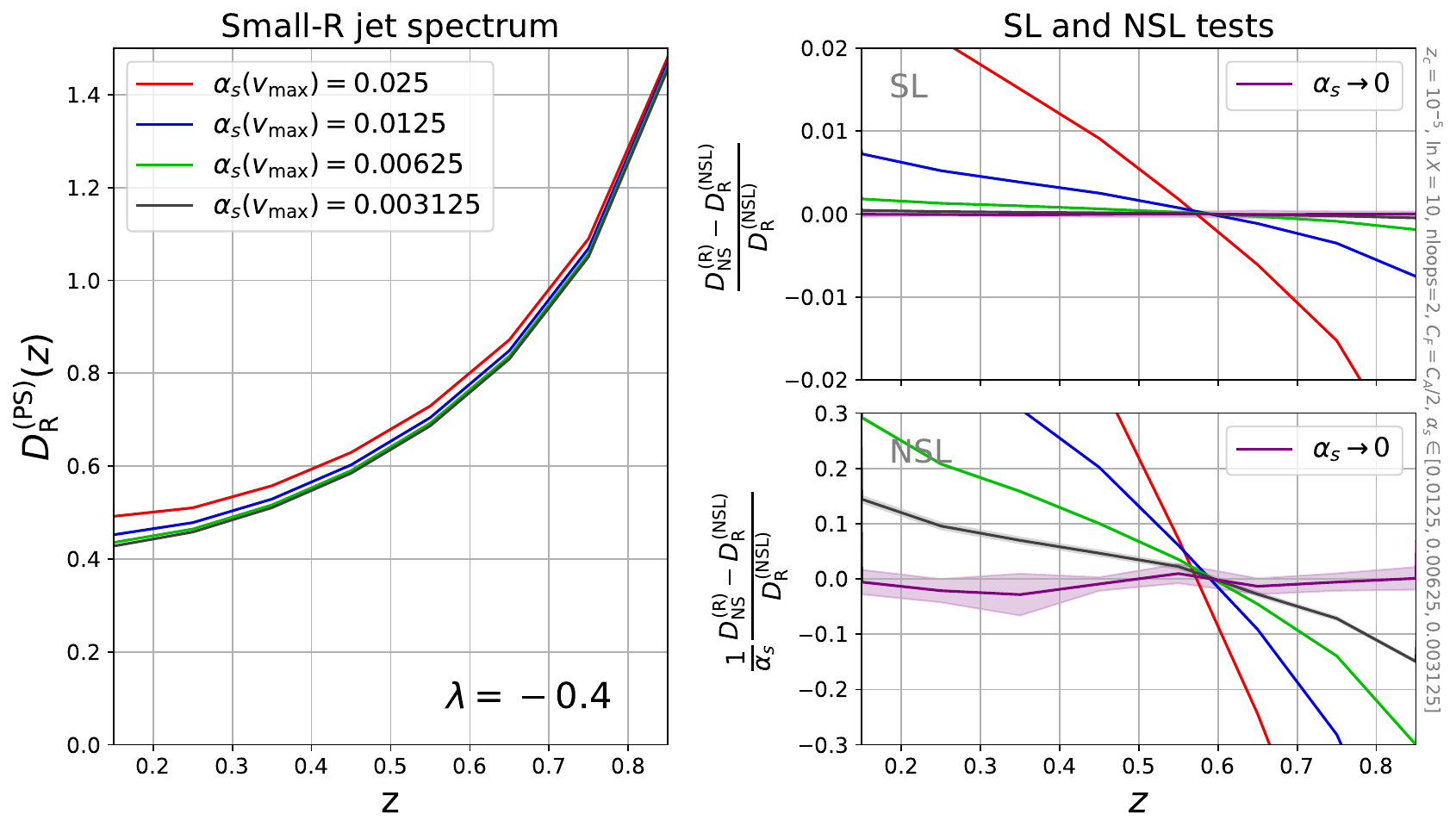}
	\caption{
	Shower small-$R$ jet spectrum and associated
	logarithmic tests at $\lambda = -0.4$, cf.\ Eq.~(\ref{eq:ratio-of-Ds-R}).
	The left-hand panel shows the non-singlet small-$R$ jet spectrum  as obtained
	with the collinear shower, for four values of the coupling.
	The upper-right panel shows a test of the shower's SL accuracy,
	where the $\as \to 0$ extrapolation gives a result consistent with
	zero, as expected for SL accuracy.
	The lower-right panel shows a test of the shower's NSL accuracy,
	illustrating that the $\as \to 0$ extrapolation is consistent with
	zero, as required for NSL accuracy.
	}
	\label{fig:smallR-test}
\end{figure}

To test the shower, we use the same procedure as for the fragmentation
function.
Fig.~\ref{fig:smallR-test} (left) shows $D_\text{R}^{(\text{PS})}(z,ER,
E R_0)$, for $\lambda = \as(E R_0) \ln R/R_0 = -0.4$ for several
values of $\as(E R_0)$.
It looks quite similar to the left-hand plot of
Fig.~\ref{fig:frag-func-test}, as is to be expected given that the
evolution of the two quantities is identical at SL level.
Fig.~\ref{fig:smallR-test} (right) shows the comparison to
Eq.~(\ref{eq:DR-reference}) as determined with HOPPET, examining the
$\as \to 0$ limit of 
\begin{equation}\label{eq:ratio-of-Ds-R}
 \frac{D^{\rm (PS)}_{\rm R}(z,ER,ER_0)}{D^{\rm
      (NSL)}_{\rm R}(z,ER,ER_0)} -1\,.
\end{equation}
for the SL test (upper panel) and the $\as \to 0$ limit of the same
quantity divided by $\as$ for the NSL test (lower panel).
The $\as \to 0$ limits are consistent with zero in both cases, as is to
be expected for an NSL-accurate collinear shower.
It is a powerful test of the shower that it reproduces NSL accuracy
both for the fragmentation function of
Section~\ref{sec:result-non-singlet-frag} and the small-$R$ jet
spectrum here.

\section{Conclusions}
\label{sec:conclusions}
In this article we have shown how to write a collinear shower that
includes the Abelian part of the real $q \to qgg$ splitting function
and the 1-loop corrections to $q \to qg$ splitting, so as to obtain
$\as^n L^{n-1}$ accuracy for specific collinear fragmentation
observables.

For the treatment of the real radiation, we made use of disordered
emissions, where strict ordering in a shower evolution variable is
replaced by strict ordering based on kinematics of the actual
post-branching partons.
This is the first use of such an approach in a study that aims for
shower logarithmic accuracy beyond SL.

Our central equation for the treatment of virtual corrections is
Eq.~(\ref{eq:K-main}), together with its simplifications in the
collinear limit. Central to our approach is
  the treatment of virtual corrections through a consistent definition
  of the inclusive NLO probability associated with the underlying
  $1\to 2$ shower splitting, as given in Eq.~\eqref{eq:K-main}.
That equation demonstrates how to account for virtual corrections as
an extension $K(z)$ of the standard soft $\Kcmw$ factor, generalised
to be differential in the $1\to2$ splitting variables, but inclusive
over subsequent branchings. Importantly, our approach can be
  applied to any shower algorithm.
Section~\ref{sec:K-determination} showed concretely how to use
Eq.~(\ref{eq:K-main}) in the collinear limit, using two independent
methods for its evaluation, one based on a slicing approach and the
other based on the approach and results of
Refs.~\cite{Dasgupta:2021hbh,vanBeekveld:2023lsa}.

Since the shower is purely collinear, its scope is limited to
observables that measure the energetic fragmentation
products of a parton, specifically non-singlet flavour combinations
(because of our inclusion, in the real part, of just $q \to qgg$
Abelian splittings).
We examined two such observables, a partonic non-singlet fragmentation
function and an inclusive small-$R$ (quark) jet momentum distribution
and confirmed that the shower reproduces known resummation results to
$\as^n L^{n-1}$ accuracy.

This work, as part of a wider goal of general NNLL accuracy for full
parton showers, provides important conceptual advances.
In particular, the recent Ref.~\cite{vanBeekveld:2024wws} was able to
achieve NNLL accuracy for event shape observables, in part by deducing
$\int_0^1 dz[K(z) - \Kcmw]$ from elements of
Ref.~\cite{Dasgupta:2021hbh,vanBeekveld:2023lsa} and specific
soft-collinear $1\to 3$ and hard-collinear $1\to2$ characteristics of
any given shower.
The understanding developed here should instead make it possible to
obtain the fully differential $K(z)$ in any given shower that includes
triple-collinear real splittings.
That will open the door to $\as^n L^{n-1}$ accuracy in full showers
also for collinear fragmentation and a wide range of jet substructure
observables.
Finally we observe that some of the methods that we have explored here
have connections with the merging of fixed-order calculations with
parton showers, notably regarding the treatment of the virtual
corrections in the approach of
Refs.~\cite{Hartgring:2013jma,Li:2016yez,Campbell:2021svd}.
As discussed recently~\cite{Hamilton:2023dwb}, for matching to
preserve logarithmic accuracy, it is critically important for the
infrared limit of matching to coincide with the shower kernels, both
in their treatment of the real and of the virtual corrections.
We believe that further exploration of such connections has the
potential to be highly fruitful for the formulation of logarithmically
accurate matching beyond NLO.

\section*{Acknowledgments}
We thank Keith Hamilton and Gregory Soyez for several discussions on
the topics of this article and for insightful comments on the
manuscript.
We are also grateful to our other PanScales collaborators (%
Silvia Ferrario Ravasio,
Alexander Karlberg,
Ludovic Scyboz,
Alba~Soto-Ontoso,
Silvia Zanoli%
)
for
various discussions on the underlying
philosophy of the approach.
We also thank each other's institutions for
their hospitality at different stages of this project.
This work has been funded by the European Research Council (ERC)
under the European Union's Horizon 2020 research and innovation
programme (grant agreement No.\ 788223, MD, BKE, JH, GPS) and under
its Horizon Europe programme (grant agreement No.\ 101044599, PM),
by a Royal Society Research Professorship
(RP$\backslash$R1$\backslash$231001, GPS) and by the U.K.'s Science
and Technologies Facilities Council under grants ST/T000864/1 (GPS),
ST/X000761/1 (GPS), ST/T001038/1 (MD) and ST/00077X/1 (MD). In the last stages of the project, BKE has been supported by the Australian Research Council via Discovery Project DP220103512.
Views and opinions expressed are however those of the authors only and
do not necessarily reflect those of the European Union or the European
Research Council Executive Agency. Neither the European Union nor the
granting authority can be held responsible for them.

\appendix

\section{Technical details}
In this appendix we discuss a number of technical aspects of the
collinear shower implementation.

\subsection{Parent-finding algorithm}
\label{sec:parent-finding}
\begin{figure}
	\centering
	\includegraphics[width=0.45\textwidth]{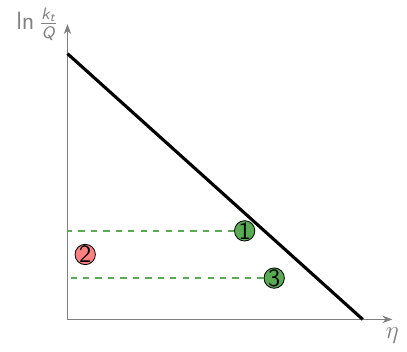}
	\caption{Lund diagram illustrating the IR safety issue that
		motivates need for a specific parent-finding algorithm.
                The fundamental issue is that when emitting $3$, the
                correct set of partons to use in the 
                triple collinear matrix element is $1$ and $3$.
                Naively using the emission immediately prior to $3$,
                i.e.\ $2$, would give an incorrect answer.
                See text for further details.
              }
	\label{fig:ir-safety}
\end{figure}

The intent of the algorithm in Section~\ref{sec:form-toy-show} is that
if two gluons are emitted commensurate in angle and in energy (and far
in angle or energy from any other gluon), then those two gluons should
be produced with the full triple collinear splitting function.
The question that we examine here is, for a given emission, how to
choose the ``parent'' for use in the triple collinear splitting
function.
An example of the issue that needs to be addressed is illustrated in
Fig.~\ref{fig:ir-safety}.
Consider the emission of gluon $i=3$ in that figure.
The situation that is dangerous is that where gluons $1$ and $3$ are
commensurate in angle and in energy (which implies $v_{1} \sim v_3$),
but there has been an intermediate emission $2$, with
$v_{2} \sim v_3$, at much lower energy (much softer) and
correspondingly at much larger angle.
The issue is that if one uses a parent $p=i-1$ ($=2$) then in the
configuration that is shown, the triple collinear correction would
be applied to the $2,3$ combination.
Because of strong angular
ordering, that correction is simply equal to $1$.
Instead, it is the $1,3$ pair that requires a non-trivial
correction, because of the proximity of $1$ and $3$ in angle.
With $p=i-1$, the presence of $2$ would prevent the triple
collinear correction from being applied to the $1,3$ pair.
The probability of $2$ being emitted between $1$ and $3$ is of the
order of $\as \ln E_q/E_{\min} \ln v_{1}/v_{3}$, where $E_{\min}$
is a cutoff on the minimal allowed radiation
energy.
This is problematic because if one imagines an actual
physical cutoff, $E_{\min} \sim k_t$, this implies
$\as \ln E_q/E_{\min}\sim 1$, i.e.\ there would be an
$\order{1}$ probability for the $1,3$ combination not
to have the correct triple-collinear correction
properly applied.
This would break NSL accuracy.
If instead one imagines taking $E_{\min}$ to be small, but with
$\ln E_q/E_{\min}$ kept finite, then we would expect NSL accuracy to
be retained, however one would have a spurious $\ln E_{\min}$ cutoff
dependence appearing at NNSL.

To avoid this problem, we use the following ``parent-finding'' algorithm.
Each emission $j$ will be associated with a variable $v_{p}(j)$.
When an emission is first created its $v_p(j)$ is set to infinity and
we also store its angle $\theta_{jq}$ with respect to the quark at its
time of creation.
When a new emission $i$ is trialled at a scale $v$, one determines the
candidate parent $p$ by identifying among all $j<i$ the one that has
the smallest value of
\begin{equation}
  \label{eq:jmin}
  \left|\ln \frac{\theta_{iq}}{\theta_{jq}}\right|\,.
\end{equation}
This is effectively the prior emission that is closest in rapidity to
$i$.
If $v_{p}(p) > v$, then we use $p$ to calculate the triple collinear
correction.
If $v_{p}(p) < v$, then we discard $i$ and continue the shower
evolution downwards from scale $v$.
In either case, all $v_{p}(j)$ for $j<i$ are reset to
$\min(v_{p}(j), v)$.
Recall that if emission $i$ is accepted, then we set
$v_{p}(i)=\infty$.

\begin{figure}
  \centering

\begin{subfigure}{0.49\textwidth}
\includegraphics[page=1,  width=\textwidth]{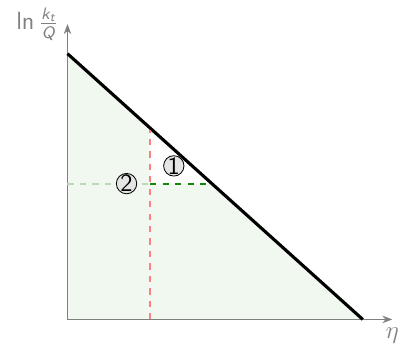}\hfill
\caption{}
\label{fig:parent-finding-algo-a}
\end{subfigure}%
\begin{subfigure}{0.49\textwidth}
\includegraphics[page=2,  width=\textwidth]{diagrams/parent_finding_algo.pdf}
\caption{}
\label{fig:parent-finding-algo-b}
\end{subfigure}

\label{fig:parent-finding-algo}
\caption{Illustration of the parent-finding algorithm, (a) for a
  simple case and (b) for a more elaborate one.
  The shaded regions indicate where emissions are allowed, with the
  vertical dashed lines identifying the boundaries of rapidity regions
  that correspond to different parent choices.
  See text for further details.  }
\end{figure}

To make the algorithm more concrete, and to help understand why it is
correct, let us examine how it functions in two example
configurations.
Fig.~\ref{fig:parent-finding-algo-a} shows a situation akin to that in
Fig.~\ref{fig:ir-safety}.
When emission $2$ was created at scale $v_2$, $v_p(2)$ was set to
$\infty$ and $v_p(1)$ to $v_2$ (shown by the horizontal dashed line).
The vertical dashed line shows the separation into rapidities that are
closer either to $2$ or to $1$.
In the unshaded region, i.e.\ when closer in rapidity to $1$ and when
$v > v_p(1)$, then an attempted emission $3$ will simply be
discarded.
If emission $3$ is closer in rapidity to $1$ but at a scale
$v < v_p(1)$, then emission $3$ will be trialled using a triple
collinear matrix-element correction that involves $1$ as a parent.
Finally if emission $3$ is closer in rapidity to $2$, then emission $3$
will be trialled with $2$ as the parent, regardless of the value of
$v$, because any value of $v$ satisfies $v < v_p(2)=\infty$.
This resolves the issue identified in Fig.~\ref{fig:ir-safety}. 
A more elaborate example is illustrated in
Fig.~\ref{fig:parent-finding-algo-b}, with the shaded area showing the
region where emission $5$ will be trialled, with the vertical lines
showing the separation into the rapidity regions that determine the
choice of parent.

\subsection{Dependence on buffer and shower cutoffs}
\label{sec:further-technical-details}
The aim of this section is to show that, for the algorithm of
Section~\ref{sec:form-toy-show} together with the parent-finding
procedure detailed in Section~\ref{sec:parent-finding}, the results
are independent (up to power corrections) of the technical parameters
of the shower algorithm.

One technical parameter is $z_c \ll 1$ and it controls the maximum and
minimum allowed gluon energies, $E_{\max} = (1-z_c)E_q$ and
$E_{\min}= z_c E_q$.
The other technical parameter is a buffer factor $X \gg 1$.
Once an emission $i-1$ has occurred at a shower evolution scale $v_{i-1}$, the shower 
continues down not from $v=v_{i-1}$ but instead from $v=X v_{i-1}$.
Note that the matrix element comes with a requirement that 
$\Theta(v_{g_i q_i} < v_{g_p q_i})$ 
(see Eq.~\eqref{eq:real-1to3}),  which together with our parent-finding
algorithm ensures that we do not double-count emissions. 
The $X \to \infty $ limit ensures that all of the
$v_{g_i q_i} < v_{g_p q_i}$ phase-space is covered.

\begin{figure}[h!]
	\centering
	\includegraphics[page=1,width=0.45\textwidth]{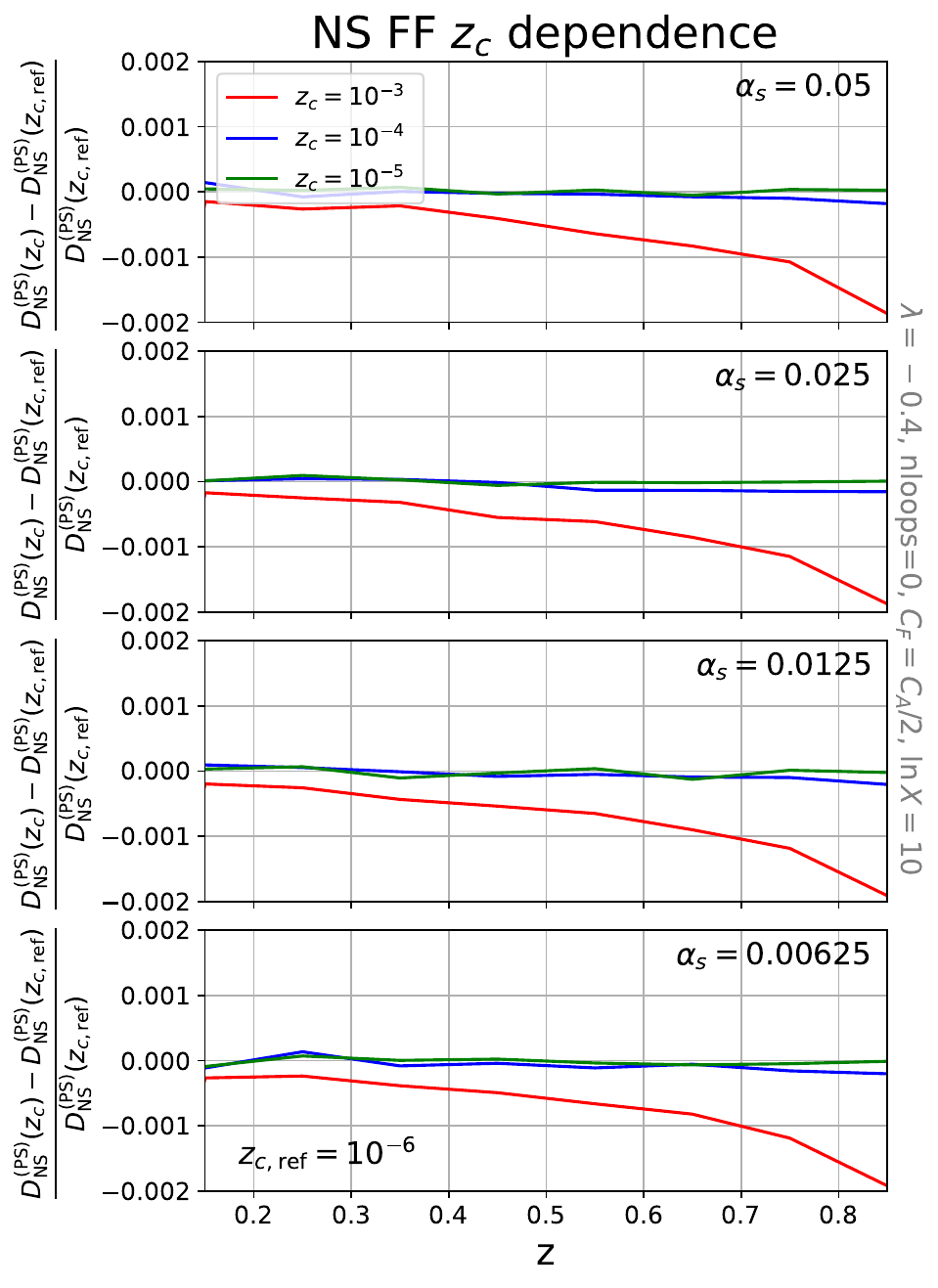}
	\includegraphics[page=2,width=0.45\textwidth]{plots/test-frag-func-zcut-buffer-dep.pdf}
	\caption{Test of $z_c$ and buffer dependence of the NS fragmentation function for fixed coupling and $\lambda = -0.4$.}
	\label{fig:zcut-buffer-dep}
\end{figure}

In Fig.~\ref{fig:zcut-buffer-dep} we show the dependence of the
spectrum of the NS fragmentation function on $z_c$ (left) and
$\ln X$ (right) for several fixed $\alpha_s$ values
through the quantity
\begin{align}
	\frac{D_{\rm NS}^{\rm (PS)}(\delta) - D_{\rm NS}^{\rm (PS)}( \delta_{\rm ref})}{D_{\rm NS}^{\rm (PS)}( \delta_{\rm ref})}\,,
\end{align}
where $\delta = z_c$ or $X$ and we have omitted the $z$,
$v_{\min}$ and $v_{\max}$ dependence of $D_{\rm NS}^{\rm (PS)}$ for
brevity.
As a reference value we take $z_c = 10^{-6}$ and $\ln X = 10$. 
We see that the result indeed tends to $0$ as long as  the energy cutoff is taken small enough ($z_c \lesssim 10^{-4}$), and the buffer size is taken large enough ($\ln X \gtrsim 3$). For all of the main results in this paper we have taken $z_c = 10^{-5}$ and $\ln X = 10$. 

\subsection{Treatment of relative angles during the shower evolution}
\label{sec:smallR-angle-treatment}

The problem that we address here is we need to apply jet clustering in
a limit where the angular separations between particles are very
small.
We work in double precision, i.e.\ numbers are stored to a relative
precision of $\epsilon \sim 10^{-16}$.
The difficulty we face is that two particles may be closer in angle
than $\epsilon$, in which case rounding errors could cause the jet
clustering to follow the wrong sequence.
To work around this difficulty we use the procedure described below.

We recall that the collinear phase-space is organised such that for
each new gluon emission $g_i$ we store its energy
$E_i \equiv E_p(1-z_i)$ (with $E_p$ being the quark energy prior to
the gluon emission), a polar angle $\theta_{g_i q_i}$ and an azimuthal
angle $\Delta \phi$.
Here $\Delta \phi$ is defined as the angle
between the plane spanned by the parent splitting with its
emitted gluon ($g_p$) and quark ($q_p$), 
and that by the decay products of that parent quark 
($q_i$ and $g_i$).
We then retrieve the energy fraction and opening angle of 
the parent splitting, $z_p$ and $\theta_{g_p q_p}$. 
The calculation of the small-$R$ jet spectrum is
sensitive to the relative angular distance between
two emissions, meaning that we need a way to store 
correctly the angular distances between all emissions. 
The collinear shower's ordering variable is $v_p =
\min[z_p,1-z_p]\theta_{g_p q_p}$, hence
smaller angle emissions do not necessarily
come after large angle emissions.
In addition, the collinear shower is not constructing actual
four-momenta, but rather only their collinear counterparts.
Therefore, care needs to be taken such that the correct angular
distances between the emissions can be obtained.
We adopt the following solution. 
Using the variables defined above, we assign
to each $1\to 3$ emission triplet an angle defined in a 2D ($x,y$)
plane
\begin{subequations}
\begin{align}
\vec{\theta}_{g_p} & = \theta_{g_p q_p}(1,0)\,, \\
\vec{\theta}_{g_i} &= -z_i\, \theta_{g_i q_i}\left(\cos \Delta \phi , \sin \Delta \phi \right)\,, \\
\vec{\theta}_{q_i} &= (1-z_i)\, \theta_{g_i q_i}\left(\cos \Delta \phi , \sin \Delta \phi \right)\,\,, \\
\theta_{g_p g_i} &= |\vec{\theta}_{g_p} - \vec{\theta}_{g_i}|\,, \quad 
\theta_{g_p q_i} = |\vec{\theta}_{g_p} - \vec{\theta}_{q_p}|\,, \quad 
\theta_{g_i q_i} = |\vec{\theta}_{g_i} - \vec{\theta}_{q_p}|\,.
\end{align}
\end{subequations}
When accepting the emission of the new $g_i$ gluon, 
we subtract $\vec{\theta}_{q_i}$ from
all emissions and from $\vec{\theta}_{q_i}$ itself.
This aligns the final-state quark with the $z$ axis with
$\vec{\theta}_{q_i} = (0,0)$. We moreover align $\vec{\theta}_{g_i}$
with the $x$ axis before the next generation step.
This makes sure that the correct relative angular
distances between the accepted gluon emissions can be obtained.

After the shower has terminated, we create massless four-momenta using
the stored energy $E_i$ and $\vec{\theta}_i$ and
\begin{align}
p_i = E_i \left(\sin \theta_i \cos \phi_i, \sin \theta_i \sin \phi_i,\cos\theta_i; 1\right), 
\end{align}
with $\theta_i = |\vec{\theta}_i|$ and
$\phi_i = \arg(\vec{\theta}_i)$.  These four-momenta are used as input
to the clustering algorithm.

\section{Evaluating $K^{\rm (ab)}_{<}(z)$}
\label{sec:kl-eval}

In this section we show how to evaluate
Eq.~\eqref{eq:Klt}, using dimensional regularisation, in the limit of emissions being collinear to the quark. We focus on the case where $i,j = g_p,g_i$, with $g_p$  having the larger $v_g$.
We use the shower phase space and map as parameterised in Eq.~\eqref{eq:mappingxtoz}.
In the limit that the slicing parameter $\lambda (\tilde{\lambda}) \ll v_g (\tilde{v}_g)$ we make use of the fact that the matrix elements and the phase space factorise. We can therefore write 
\begin{subequations}
\begin{multline}
	\label{eq:phase-space-me-3}
\frac{d\Phi_{q\bar q g_1 g_2}}{d\Phi_{q\bar q g}}\frac{B_{q\bar q g_1 g_2}}{B_{q \bar q g}}
 = \frac{e^{\epsilon\gamma_\mathrm{E}}\Gamma(1-\epsilon)}{\Gamma(1-2\epsilon)} 
 P_{qq}(z_i,\epsilon)\,
\delta\left(\left(\frac{v_g}{E \min(z_{p},1-z_{p})}\right)^2- \, \theta^2_{g_pq_p}\right) \,
 \\ 
\times \delta(z-z_p) (z_p z_i(1-z_i))^{-2\epsilon}
\sin \Delta \phi^{-2\epsilon}
dz_p dz_i\, d\theta_{g_p q_p}^2\, \frac{d\theta_{g_i q_i}^2}{\theta^{2+2\epsilon}_{g_i q_i}}\,  \frac{d\Delta \phi}{2\pi}
\end{multline}
and 
\begin{equation}
	\label{eq:phase-space-me-2}
\frac{d\Phi_{q\bar q g'}}{d\Phi_{q\bar q}}\frac{B_{q\bar q g'}}{B_{q \bar q}} = 
\frac{e^{\epsilon\gamma_\mathrm{E}}}{\Gamma(1-\epsilon)}
P_{qq}(z_p,\epsilon)(z_p(1-z_p))^{-2\epsilon} 
dz_p\, \frac{d\theta_{g'q}^2}{ \theta_{g'q}^{2+2\epsilon} }\,.
\end{equation}
\end{subequations}
Here we have used
\begin{equation}
P_{qq}(z,\epsilon) = C_F\left(\frac{1+z^2}{1-z}-\epsilon(1-z)\right)\,.
\end{equation}
These are to be integrated in $v_{g_i q_i} = [0,\lambda]$ and
$v_{g'q} = [0,\lambda]$ respectively (see Eqs.~\eqref{eq:Klt},\eqref{eq:lambda-conditions}), recalling that we take $\tilde{\lambda}=\lambda$ as per section~\ref{sec:K-CF-slicing}. 
We decide to write these phase-space constraints as 
\begin{subequations}
\begin{align}
\Theta(\lambda > \min(E_{g_i}, E_{q_i}) \theta_{g_i q_i}) &=
1-\Theta(\lambda/E < z_p\min(z_i, 1-z_i ) \theta_{g_i q_i})\,, \\
\Theta(\lambda > \min(E_{g'}, E_q)  \theta_{g' q})
&= 1-\Theta(\lambda/E < \min(z_{p}, 1-z_{p})  \theta_{g' q})\,,
\end{align}
\end{subequations}
such that there is one integral to be carried out in $4-2\epsilon$ dimensions, while the two other integrals involving a phase space constraint are finite in $4$ dimensions. 
We then perform the integrals over $\Delta \phi$ and $\theta_{g_pq_p}$ in Eq.~\eqref{eq:phase-space-me-3} and perform a change-of-variables
to $z_i, \theta_{g_i q_i} \to z',\theta'$ for Eq.~\eqref{eq:phase-space-me-3} , 
and $z_p, \theta_{g' q} \to z',\theta'$  for Eq.~\eqref{eq:phase-space-me-2}
so as to bring the two integrands in a common form.
After this 
the two real contributions of Eq.~\eqref{eq:Klt} can be written as
\begin{align}
\label{eq:Klreal_eval}
\int^{\tilde \lambda}_0
               \frac{d\Phi_{q\bar q ij}}{d\Phi_{q\bar q g}}
               \frac{B_{q\bar q ij}}{B_{q \bar q g}}
               - \int^{\lambda}_{0}
               \frac{d\Phi_{q\bar q g'}}{d\Phi_{q\bar q}}
               \frac{B_{q\bar q g'}}{B_{q \bar q}}& = \\
&\hspace{-4cm}+ \int_0^{\theta_{\max}^2}\frac{d\theta'^{2}}{\theta'^{2(1+\epsilon)}} \int_0^1 dz' P_{qq}(z',\epsilon) z^{\prime-2\epsilon}(1-z^{\prime})^{-2\epsilon}
\left( z^{-2\epsilon}-1\right) \nonumber
 \\
&\hspace{-4cm}- \int_0^{\theta_{\max}^2} \frac{d\theta'^{2}}{\theta'^2} \int_0^1 dz' P_{qq}(z')  
\Theta\left(z \min(z',1-z') \theta' > \lambda/E \right) \nonumber\\
&\hspace{-4cm}+\int_0^{\theta_{\max}^2} \frac{d\theta'^{2}}{\theta'^2} \int_0^1 dz' P_{qq}(z') 
\Theta\left(\min(z',1-z') \theta' > \lambda/E \right)\,. \nonumber
\end{align}
The factor of $z^{-2\epsilon}-1$ on the second line originates in the difference between the three and two particle phase spaces in $d$ dimensions, and gives rise to a single pole which cancels against a corresponding pole in the virtual corrections.
The constraint on the maximum angle $\theta_{\rm max}$ ensures
that we stay in the collinear regime.

The integrals in Eq.~\eqref{eq:Klreal_eval} can be readily evaluated, and read
\begin{align}
	\label{eq:reals-eval}
	\int^{\tilde \lambda}_0
	\frac{d\Phi_{q\bar q ij}}{d\Phi_{q\bar q g}}
	\frac{B_{q\bar q ij}}{B_{q \bar q g}}
	- \int^{\lambda}_{0}
	\frac{d\Phi_{q\bar q g'}}{d\Phi_{q\bar q}}
	\frac{B_{q\bar q g'}}{B_{q \bar q}}& = \\
& \hspace{-2cm}	- 2C_F\ln(z)\left(\frac{1}{\epsilon}-\ln\frac{\lambda^2}{E^2}\right)\,. \nonumber
\end{align}
The virtual corrections are (up to $\mathcal{O}(\epsilon)$ terms)
\begin{align}
	\label{eq:virt-corr}
\frac{V_{q\bar q g}}{B_{q\bar q g}}
               - \frac{V_{q\bar q}}{B_{q\bar q}} &=  \\
& \hspace{-1cm} 2C_F
 \left(\frac{\ln (z)}{\epsilon } +  \mathrm{Li}_2\left(\frac{z-1}{z}\right) - \ln \frac{s_{qg}}{E^2}\ln (z) \right)-\frac{C_F}{p_{qq}(z)}\,, \nonumber
\end{align}
where we use $s_{qg} = z(1-z)v_g^2/\min(z,1-z)^2$.
Together with Eq.~\eqref{eq:reals-eval} we find the result given in Eq.~\eqref{eq:Klt-result}.

\section{Definition of ${\cal B}_2^q(x)$}
\label{sec:B2}
In this appendix we outline the expression of the quantity
${\cal B}_2^f(z)$ defined in
Refs.~\cite{Dasgupta:2021hbh,vanBeekveld:2023lsa}.
In particular, we limit ourselves to reporting the expressions for the
quark case, ${\cal B}_2^q(z)$, since the present article focusses on
the non-singlet flavour channel.

In the notation of Ref.~\cite{vanBeekveld:2023lsa}, ${\cal B}_2^q(z)$
can be expressed as~\footnote{The function ${\mathcal B}^q_2(z)$
  computed in Ref.~\cite{Dasgupta:2021hbh} is defined as the
  ${\mathcal B}^q_2(z)$ used here multiplied by
  $\left(\alpha_s/(2\pi)\right)^2$.}
\begin{equation}\label{eq:B2def}
{\mathcal B}^q_{2}(z) = {\mathcal B}^{q,\,{\rm an.}}_{2}(z) + C_F^2  H^{\rm fin.}(z)\,,
\end{equation}
where
\begin{align}\label{eq:B2q}
	 {\mathcal B}^{q,\,{\rm an.}}_{2}(z) &=
                                       C^2_F\,
  \mathcal{B}_{2}^{q,{\rm an.},\ab}(z) +C_F C_A\, \mathcal{B}_{2}^{q,{\rm an.},\,C_F C_A}(z) +C_F T_R n_f \,
  \mathcal{B}_{2}^{q,{\rm an.},\,C_F T_R}(z) \notag\\&+ C_F\left(C_F - \frac{C_A}{2}\right)
  \mathcal{B}_{2}^{q,{\rm an.},\textrm{id.}}(z)  \ .
\end{align}
The term $ \mathcal{B}_{2}^{q,{\rm an.},\textrm{id.}}(z) $ is
neglected in the results presented in this article, which adopt the
leading-$N_c$ limit, i.e.\ $C_F=C_A/2$.
To simplify the notation we absorb the term
${\mathcal B}^q_{1}(z) b_0\ln g^2(z)$ in Eq.~(2.10)
of~\cite{vanBeekveld:2023lsa} into ${\mathcal B}^q_{2}(z)$.
  In contrast with the convention adopted
  in~\cite{vanBeekveld:2023lsa}, where we set $g(z)=1-z$, this term
  multiplies now the whole splitting function to allow for the
  possibility to set $g(z)\neq 1-z$ also in the soft term of the
  inclusive emission probability. In practice, we use $g(z)=1-z$ for
  the angular ordering case while the ordering variable defined in
  Eq.~\eqref{eq:vi-definition} corresponds to using
  $g(z) = \min(z,1-z)$ in the scale of the coupling.
  The functions in Eq.~\eqref{eq:B2q} read
\begin{align}\label{eq:B2-expressions}
	\mathcal{B}_{2}^{q,{\rm an.},\,\textrm{id.}} (z) &= 4z-\frac{7}{2} +\frac{5z^2-2}{2(1-z)} \ln z+\frac{1+z^2}{1-z}\left(\frac{\pi^2}{6} -\ln z \ln (1-z)-\text{Li}_2(z)\right)\,, \\
	\mathcal{B}_{2}^{q,{\rm an.},\,C_F T_R} (z) &= - b_0^{(n_f)} p_{qq}(z) \ln z + b_0^{(n_f)} (1-z) - K^{(1),n_f} (1+z) \notag\\&\quad+ 2\,b_0^{(n_f)} (1+z) \ln(1-z)+b_0^{(n_f)} p_{qq}(z) \ln(g^2(z))\, , \\ 
	\mathcal{B}_{2}^{q,{\rm an.},\,C_F C_A}(z) &= - b_0^{(C_A)} p_{qq}(z) \ln z + b_0^{(C_A)} (1-z) +\frac{3}{2} \frac{z^2 \ln z}{1-z} +\frac12 (2z-1)\\& + p_{qq}(z) \left(\ln^2z+
   \mathrm{Li}_2\left(\frac{z-1}{z}\right)+2\,
   \mathrm{Li}_2(1-z)\right) - K^{(1),C_A} (1+z) \notag\\
&+2\,b_0^{(C_A)} (1+z) \ln (1-z)+b_0^{(C_A)} p_{qq}(z) \ln(g^2(z))\,,\notag\\ \label{eq:B2qCFCF}
  \mathcal{B}_{2}^{q,{\rm an.},\ab} (z) &= p_{qq}(z) \left(-3 \ln z -2 \ln z \ln(1-z) +2\, \text{Li}_2\left(\frac{z-1}{z}\right)\right)  - 1 \,,
\end{align}
where we have used the decomposition
\begin{equation}
K^{(1)}=\left(\frac{67}{18} - \frac{\pi^2}{6}\right)\,C_A -
\frac{10}{9}\,T_R\,n_f \equiv K^{(1),C_A}\,C_A + K^{(1),n_f}\,T_R\,n_f \,,
\end{equation}
for the two-loop coefficient of the physical coupling scheme and 
\begin{align}\label{eq:b0}
  b_0 = \frac{11}{6}\, C_A - \frac23 \,T_R \,n_f \equiv b_0^{(C_A)} \, C_A + b_0^{(n_f)}\, T_R \,n_f \,,
\end{align}
for the first coefficient of the QCD beta function $b_0$.
The function $g(z)$ parameterises the scale of the running coupling
used in the Sudakov factor.
A crucial aspect of Eq.~\eqref{eq:B2def} is that the quantity
$ {\mathcal B}^{q,\,{\rm an.}}_{2}(z) $ is independent of the
ordering variable, and the complete dependence on the ordering is
encoded in $H^{\rm fin.}(z)$.

For the case of the ordering variable~\eqref{eq:vi-definition} this is
defined in Eq.~\eqref{eq:Hfin-v}, while in the angular-ordered case it
is given by the integral
\begin{align}\label{eq:Hfin-theta}
  H_{\theta}^{\rm fin.}(z) &= \int
  d\Phi_{3}(8 \pi \alpha_s)^2 \left(\frac{\langle
  P\rangle_{C_F^2}}{s_{g_p g_i q_i}^2} - \left[\frac{\langle P\rangle_{C_F^2}}{s_{g_p g_i
        q_i}^2}\right]_{\theta_{g_p q_i}\gg \theta_{g_i q_i}} \right) \,\Theta(\theta_{g_p q_i} - \theta_{g_i q_i})\notag\\&\times\delta(z - z_p)\,\theta^2\delta(\theta^2-\theta_{g_pq_i}^2)\,,
\end{align}
where we adopt the notation of Eqs.~\eqref{eq:B-splits},~\eqref{eq:Pcf2},~\eqref{eq:JPS}.
The invariant mass of the three-particle collinear system is denoted
by $s_{g_p g_i q_i}$. The spin averaged $1\to 3$ splitting function
$\langle P\rangle_{C_F^2}=\langle
P\rangle_{C_F^2}(x_{g_p},x_{g_i},x_{q_i})$ (with $x_i$ being the
longitudinal momentum fraction of particle $i$ w.r.t.~the energy of
the initiating parton) can be found in Ref.~\cite{Catani:1998nv}, and
$\left[\frac{\langle P\rangle_{C_F^2}}{s_{g_p g_i
      q_i}^2}\right]_{\theta_{g_p q_i}\gg \theta_{g_i q_i}}$ denotes
the limit of the quantity
$\frac{\langle P\rangle_{C_F^2}}{s_{g_p g_i q_i}^2} $ in the strongly
ordered regime $\theta_{g_p q_i}\gg \theta_{g_i q_i}$, and it is given
in Eq.~\eqref{eq:Pcf2}.
Finally, the three-body phase space measure $d\Phi_{3}$, including the
Jacobian factor for the chosen parametrisation, is given by ($\Delta$
denotes the Gram determinant, see
Refs.~\cite{Dasgupta:2021hbh,vanBeekveld:2023lsa})
\begin{align}\label{eq:psB}
  \sd \Phi_3\equiv\frac{z_p}{\pi}
  \frac{E^{4}}{(4\pi)^{4}} \sd
  z_p \,\sd z_i\, \sd \theta_{g_p q_i}^2\, \sd  \theta_{g_i q_i}^2\, \sd \theta_{g_p g_i}^2\,
  (1-z_p)\, z_p^2\, (1-z_i)\, z_i \,\Delta^{-1/2} \,\Theta(\Delta) \,.
\end{align}
Eq.~\eqref{eq:Hfin-theta} can be reduced to a 1-fold integral (cf.\ Fig.~4 
of Ref.~\cite{Dasgupta:2021hbh}), that is provided as
an ancillary file with Ref.~\cite{vanBeekveld:2023lsa}.
The function $ {\mathcal B}^q_{2,\theta}(z)$ is regular in the soft limit
$z \to 1$ and is thus fully integrable over $z \in [0,1]$.

\bibliographystyle{JHEP}
\bibliography{triple-collinear}
\end{document}